\documentclass[twocolumn]{aastex631}

\newcommand{\makechange}[1]{\textcolor{blue}{\textbf{#1}}}

\usepackage{float}
\usepackage[utf8]{inputenc}
\usepackage{textgreek}
\usepackage{booktabs}
\usepackage{appendix}

\begin{document}

\title{Environmental Imprints on the Assembly of the Cool Gas around Bright Cluster Galaxies}

\correspondingauthor{Hu Zou, Taotao Fang}
\email{zouhu@nao.cas.cn, fangt@xmu.edu.cn}

%


\author[0000-0002-8037-3573]{Runyu Zhu}
\affiliation{Department of Astronomy, Xiamen University, Xiamen, Fujian 361005, People’s Republic of China} 
\affiliation{National Astronomical Observatories, Chinese Academy of
Sciences, Beijing 100012, People’s Republic of China}

\author[0000-0002-6684-3997]{Hu Zou}
\affiliation{National Astronomical Observatories, Chinese Academy of Sciences, Beijing 100012, People’s Republic of China} 
\affiliation{School of Astronomy and Space Science, University of Chinese Academy of Sciences, Beijing 101408, People’s Republic of China}

\author{Taotao Fang}
\affiliation{Department of Astronomy, Xiamen University, Xiamen, Fujian 361005, People’s Republic of China}

\author{Enci Wang}
\affiliation{Department of Astronomy, University of Science and Technology of China, Hefei, Anhui 230026, People’s Republic of China \\}

\affiliation{School of Astronomy and Space Science, University of Science and Technology of China, Hefei, Anhui 230026, People’s Republic of China \\}

\author{Zeyu Chen}
\affiliation{Department of Astronomy, University of Science and Technology of China, Hefei, Anhui 230026, People’s Republic of China \\}
\affiliation{School of Astronomy and Space Science, University of Science and Technology of China, Hefei, Anhui 230026, People’s Republic of China \\}

\author{Zhaobin Chen}
\affiliation{School of Physics and Technology, Nanjing, Jiangsu 210023, People’s Republic of China \\}
\affiliation{National Astronomical Observatories, Chinese Academy of Sciences, Beijing 100012, People’s Republic of China \\} 

\author{Weiyu Ding}
\affiliation{Department of Astronomy, University of Science and Technology of China, Hefei, Anhui 230026, People’s Republic of China \\}
\affiliation{School of Astronomy and Space Science, University of Science and Technology of China, Hefei, Anhui 230026, People’s Republic of China \\}

\author{Jinfu Gou}
\affiliation{National Astronomical Observatories, Chinese Academy of
Sciences, Beijing 100012, People’s Republic of China}
\affiliation{School of Astronomy and Space Science, University of Chinese Academy of Sciences, Beijing 101408, People’s Republic of China}

\author{Weijian Guo}
\affiliation{National Astronomical Observatories, Chinese Academy of
Sciences, Beijing 100012, People’s Republic of China}

\author{Niu Li}
\affiliation{National Astronomical Observatories, Chinese Academy of
Sciences, Beijing 100012, People’s Republic of China}

\author{Wenxiong Li}
\affiliation{National Astronomical Observatories, Chinese Academy of
Sciences, Beijing 100012, People’s Republic of China}

\author{Shufei Liu}
\affiliation{National Astronomical Observatories, Chinese Academy of
Sciences, Beijing 100012, People’s Republic of China}
\affiliation{School of Astronomy and Space Science, University of Chinese Academy of Sciences, Beijing 101408, People’s Republic of China}

\author{Haoming Song}
\affiliation{National Astronomical Observatories, Chinese Academy of
Sciences, Beijing 100012, People’s Republic of China}
\affiliation{School of Astronomy and Space Science, University of Chinese Academy of Sciences, Beijing 101408, People’s Republic of China}

\author{Jipeng Sui}
\affiliation{National Astronomical Observatories, Chinese Academy of
Sciences, Beijing 100012, People’s Republic of China}
\affiliation{School of Astronomy and Space Science, University of Chinese Academy of Sciences, Beijing 101408, People’s Republic of China}

\author{Xi Tan}
\affiliation{National Astronomical Observatories, Chinese Academy of
Sciences, Beijing 100012, People’s Republic of China}
\affiliation{School of Astronomy and Space Science, University of Chinese Academy of Sciences, Beijing 101408, People’s Republic of China}

\author{Yunao Xiao}
\affiliation{National Astronomical Observatories, Chinese Academy of
Sciences, Beijing 100012, People’s Republic of China}
\affiliation{School of Astronomy and Space Science, University of Chinese Academy of Sciences, Beijing 101408, People’s Republic of China}

\author{Jingyi Zhang}
\affiliation{National Astronomical Observatories, Chinese Academy of
Sciences, Beijing 100012, People’s Republic of China}



\begin{abstract}
Galaxy clusters represent extreme cosmic laboratories where environmental processes dramatically reshape their constituent galaxies, yet their effect on the gaseous halos of central galaxies remains poorly constrained. Here we present the first statistical mapping of cool gas around massive brightest cluster galaxies (BCGs) at $z\approx0.55$. Using Mg II absorption in stacked sight-line spectra from over a million background quasars observed by the Dark Energy Spectroscopic Instrument, we compare BCGs to a matched sample of field galaxies and trace the radial profile from 40 kpc to 15 Mpc. Our analysis reveals a striking dual environmental signature: within 200 kpc, the circumgalactic medium (CGM) around BCGs is significantly suppressed compared to that of field galaxies, while at larger radii (200 kpc to 10 Mpc) a pronounced excess of cool gas emerges. This clear transition from suppression in the core to enhancement on such large scales delineates a novel observed pattern for gas regulation by the dense environment. It suggests that clusters may not only strip gas in the core but also facilitate its accumulation in the outskirts. Our results provide key observational constraints on theoretical models of environmental processing in and around the most massive dark matter halos.
\end{abstract}



\section{Introduction} \label{sec:intro}

Galaxy clusters are the most massive gravitationally bound systems in the universe and serve as critical laboratories for studying galaxy evolution in extreme environments. Within their deep potentials, a hot intracluster medium (ICM) and strong tidal fields drive environmental processes such as ram-pressure stripping, tidal stripping, and starvation \citep{1972ApJ...176....1G, 2022A&ARv..30....3B, 1990ApJ...350...89B, 1980ApJ...237..692L}. These mechanisms efficiently remove gas from infalling galaxies, resulting in a population dominated by quiescent, early-type systems \citep{1984ApJ...281...95P, 2010ApJ...721..193P}. Intriguingly, many of these quiescent galaxies are observed to harbour substantial reservoirs of cool, metal-enriched gas in their halos \citep[e.g.,][]{Lundgren_2009ApJ...698..819L, Huang_2016MNRAS.455.1713H, Zahedy_2019MNRAS.484.2257Z}. The origin and spatial distribution of this cool gas remain active topics of research.

The evolution of galaxies is fundamentally tied to the circumgalactic medium (CGM), encompassing the multiphase gas within their dark matter halos that regulates the baryon cycle through accretion and outflows \citep{Tumlinson_2017ARA&A..55..389T, 2020ARA&A..58..363P}. While internal feedback processes, such as supernovae and active galactic nuclei, significantly influence the CGM \citep[e.g.,][]{Weinberger_2017MNRAS.465.3291W, Burkhart_2022ApJ...933L..46B}, the external cluster environment is increasingly recognized as a dominant influence. Some studies suggesting that the cool CGM around luminous red galaxies (LRGs) may correlate more closely with the extended dark matter halo than with the central galaxy itself \citep{Smailagic_2018ApJ...867..106S}. A key unanswered question is how cluster environments reshape the gaseous halos of their central galaxies: does the dense environment primarily suppress and destroy cool CGM gas, or does it actively redistribute it, potentially enhancing gas reservoirs at larger scales? 

Brightest cluster galaxies (BCGs) are central to addressing this question, as they reside at the dynamical center of clusters and their evolution is coupled to the growth of the cluster
residing at the dynamical centers of clusters \citep{1977ApJ...212..311T, 2007MNRAS.379..867V, 2009MNRAS.398.1698S, 2024ApJ...963...21S}. Recent cosmological magnetohydrodynamical simulations provide a predictive framework, suggesting that cool gas in clusters is not distributed uniformly but is preferentially located in the outskirts, associated with infalling structures \citep{2025A&A...703A..33S}. Nevertheless, direct observational evidence of how the cluster environment affects the CGM and larger-scale intergalactic medium (IGM) of BCGs remains very limited.

The Mg II $\lambda\lambda$2796, 2803 doublet serves as an excellent tracer of cool gas \citep[$T\sim10^{4}$ K;][]{Charlton_2003ApJ...589..111C}. Its utility for probing metal-enriched, cool gaseous structures was demonstrated in the pioneering study of \citet{1986A&A...169....1B}, who detected the doublet in absorption toward a galaxy pair. Subsequently, many works have used Mg II absorption lines in the spectra of bright background sources, such as quasars, to study the cool CGM of foreground galaxies (e.g., \citealp{2014MNRAS.439.3139Z, Lan2018ApJ, 2020MNRAS.499.5022D, Anand_2021}; \citealp{2024A&A...684A.136F}; \citealp{Wu_2025ApJ...983..186W, 2025ApJ...981...81C, 2025arXiv251000729U, 2025OJAp....847836H, 2025arXiv251203845C})
Nevertheless, the influence of the cluster environment on the CGM of BCGs remains poorly constrained. The advent of the Dark Energy Spectroscopic Instrument \citep[DESI;][]{PART1, PART2} provides an unprecedented opportunity to address these questions. With its vast spectroscopic sample of background quasars, DESI enables statistical mapping of Mg II absorption around foreground galaxies with unprecedented precision, from the inner CGM (tens of kpc) out to the large-scale IGM (tens of Mpc). 

Building upon the critical role of cluster environments in shaping galaxy evolution, this study aims to obtain critical observational evidence on how these extreme conditions affect the cool gas halo around BCGs. We employ the unprecedented spectroscopic sample from the DESI DR1 data to statistically map Mg II absorption from the inner CGM out to the large-scale IGM. This is achieved by stacking background quasar spectra aligned with foreground galaxies. We systematically compare the absorption profiles around BCGs with those around a carefully matched sample of field LRGs, controlling for redshift and stellar mass to isolate the dense environmental impact. Here, the term “field” is defined relative to the BCGs in the clusters used in our study, and refers collectively to environments that are significantly less dense than the cluster regime; it includes both isolated field galaxies and those in galaxy groups, where the gravitational potential and matter density are substantially lower than in the central regions of massive clusters.

This Letter is organized as follows: Section \ref{sec:data} details the dataset and sample selection, Section \ref{sec:methods} outlines our data processing and spectral stacking methodology, and Section \ref{sec:results} presents our main findings. We summarize in Section \ref{sec:summary}. We assume a flat ΛCDM cosmology with Planck 2018 parameters: $h=0.677$, $\Omega_{\mathrm{M}}=0.310$, and $\Omega_{\mathrm{\Lambda}}=0.688$ \citep{Planck18}.

\section{Data and Sample} \label{sec:data}

We investigate the properties of the CGM and IGM by statistically analyzing the Mg II $\lambda\lambda2796, 2803$ doublet absorption features in the spectra of background quasars. These absorption lines arise when the sight-line to a quasar passes through cool gas structures at different projected distances (i.e., impact parameters) from foreground galaxies. Although individual absorption features are often too weak to detect, we employ spectral stacking techniques (Section~\ref{sec:methods}) to extract robust average signals from both the CGM and the large-scale IGM. To isolate the influence of environment, we adopt a comparative approach using two distinct galaxy populations: BCGs and field massive LRGs. 

\subsection{DESI Data} \label{subsec:data-sets}

Both the background quasars and the foreground galaxies are selected from the DESI DR1 catalog, specifically the file \texttt{zall-pix-iron.fits} \citep{DESI_DR1}. This catalog contains nearly 30 million sources, incorporating data from the DESI Commissioning, Survey Validation, Main Survey, and special programs observed between December 14, 2020 and June 13, 2022. The DR1 dataset increases the spectral coverage by an order of magnitude compared to the Early Data Release (EDR; \citealt{EDR}), enabling a statistically robust investigation of the distribution of cool gas using a large sample of both foreground and background sources with wide sky coverage.

Spectroscopic redshifts for all sources are determined by the \texttt{Redrock} data analysis pipeline \footnote{\url{https://github.com/desihub/redrock}} (S.Bailey et al. 2024, in prep), which performs template fitting to classify objects and measure their redshifts. We include only sources with reliable spectroscopic redshifts, as indicated by the flag \texttt{ZWARN $==$ 0}. The spectral classification is given in the \texttt{SPECTYPE} column. We identify background quasar sight-lines by selecting objects with \texttt{SPECTYPE $==$ "QSO"}. Foreground galaxies are identified using \texttt{DESI\_TARGET} bitmask \footnote{\url{https://github.com/desihub/desitarget/blob/main/py/desitarget/data/targetmask.yaml}}, a 64-bit integer where each bit corresponds to a specific target class \citep{TARGET_2N}. The bit 0 identifies LRGs. Given the DESI wavelength coverage (3600--9800 \AA), we impose a redshift cut of $z>0.4$ on foreground galaxies to ensure that the Mg II doublet falls within the observed spectral range. To incorporate key physical properties into our analysis, we obtain the stellar masses of our galaxy sample from the DESI DR1 Stellar Mass and Emission Line value-added catalog (hereafter the \textit{mass} catalog) \footnote{\url{https://data.desi.lbl.gov/doc/releases/dr1/vac/stellar-mass-emline/}} \citep{2024ApJ...961..173Z}.

\subsection{BCG Sample} \label{subsec:BCG}

We select BCG sample from a galaxy cluster catalog \citep{BCG_catalog} based on the DESI Legacy Imaging Surveys Data Release 9 (DR9; \citealt{2019AJ....157..168D}), which contains over 530,000 clusters. To obtain reliable redshifts, we cross-match this sample with the DESI DR1 catalog. For matched BCGs, we adopt the DESI spectroscopic redshift and stellar mass from the \textit{mass} catalog; for unmatched BCGs, we retain existing high-quality spectroscopic redshifts from the cluster catalog. After applying a redshift cut of $z>0.4$ to ensure Mg II coverage, our final sample comprises 103,809 BCGs, with a median redshift of $z = 0.54$ and median stellar mass of $\mathrm{log}(M_{*}/M_{\odot}) = 11.46$.

\subsection{Control sample of field massive galaxies} \label{subsec:fields}

To investigate environmental effects on the cool gas halos around massive galaxies, we construct a control sample of field galaxies that are intrinsically similar to BCGs (e.g., the stellar mass) but reside in lower-density environments. We select LRGs from the DESI DR1 catalog \citep{2023JCAP...11..097Z}. LRGs are ideal as they are massive systems similar to BCGs. To minimize contamination from cluster members, we apply the following criteria: (1) reliable spectroscopic redshift (\texttt{ZWARN == 0}) and $z>0.4$; (2) exclusion if within 15 Mpc of any BCG to avoid nearby cluster contamination; and (3) inclusion only if the nearest BCG is within 50 Mpc to maintain consistent sky coverage with the BCG sample.

\begin{figure}[ht!]
\hspace*{-1\columnsep}
\includegraphics[width=1.2\linewidth,height=1.2\textheight,keepaspectratio, trim=0 20pt 0 80pt, clip]{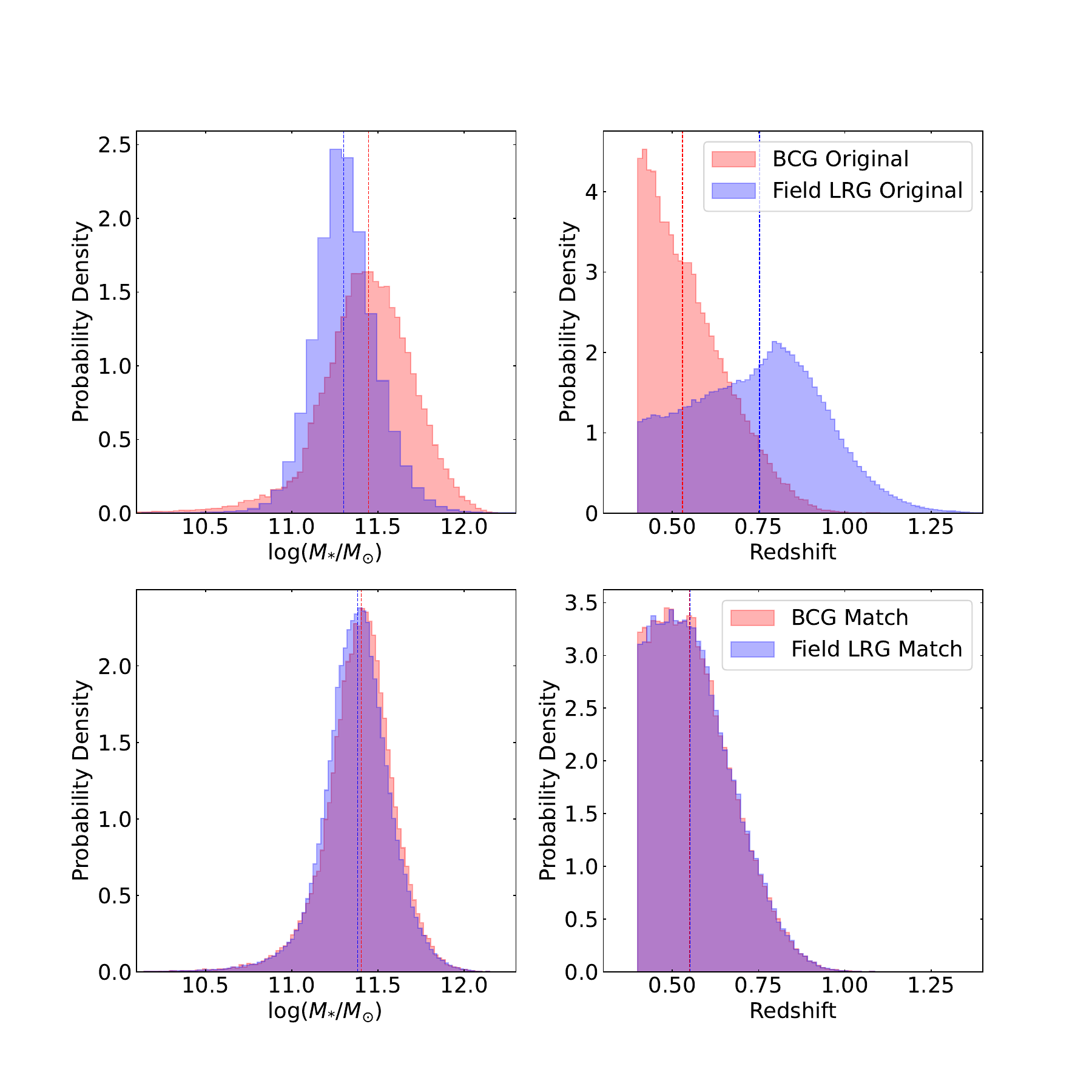}
\caption{
Probability density distributions of stellar mass (left column) and redshift (right column) for the BCG and LRG samples. Vertical dashed lines indicate median values, with colors corresponding to the legend. Top two panels: Original samples selected according to Sections~\ref{subsec:BCG} and~\ref{subsec:fields}. The BCG sample includes only those with at least one background quasar, as BCGs without sight-lines do not contribute to stacking. Bottom two panels: Matched subsamples after unifying the redshift and stellar mass distributions.
\label{fig:z-mass}}
\end{figure}

As shown in the top panels of Figure \ref{fig:z-mass}, BCGs in the DESI original sample are systematically biased toward lower redshifts and higher stellar masses compared to field LRGs. To address this, we performed a distribution-matching procedure in the redshift–mass plane to enable a fair comparison. Using the BCGs as a reference sample, we selected corresponding field LRGs that satisfied the matching criteria: $|z_{\rm{LRG}} - z_{\rm{BCG}}| < 0.005$ and $|\rm{log}(\mathit{M}_{\rm{LRG}}/\mathit{M}_{\rm{\odot}}) - \rm{log}(\mathit{M}_{\rm{LRG}}/\mathit{M}_{\rm{\odot}})| < 0.03$. We allow a maximum of five LRG matches per BCG and exclude any BCG with fewer than one match, ensuring a balanced comparison while using each LRG only once. Furthermore, we included only those BCGs that had at least one background quasar in the sight-line, as BCGs without background quasars do not contribute to the stacked absorption signal. The resulting matched samples (shown in the bottom panels of Figure~\ref{fig:z-mass}) exhibit closely aligned distributions in both redshift and stellar mass. The final matched sample consists of 72,913 BCGs and 364,565 field LRGs. The samples have a median redshift of $z\approx0.55$, with 90\% of galaxies located at $z<0.72$. Within this redshift range, our galaxy cluster sample is complete. By confining our analysis to this relatively narrow redshift range, we largely mitigate the potential impact of redshift evolution.

\begin{figure}[ht!]
\centering
\includegraphics[width=1\linewidth,height=1\textheight,keepaspectratio]{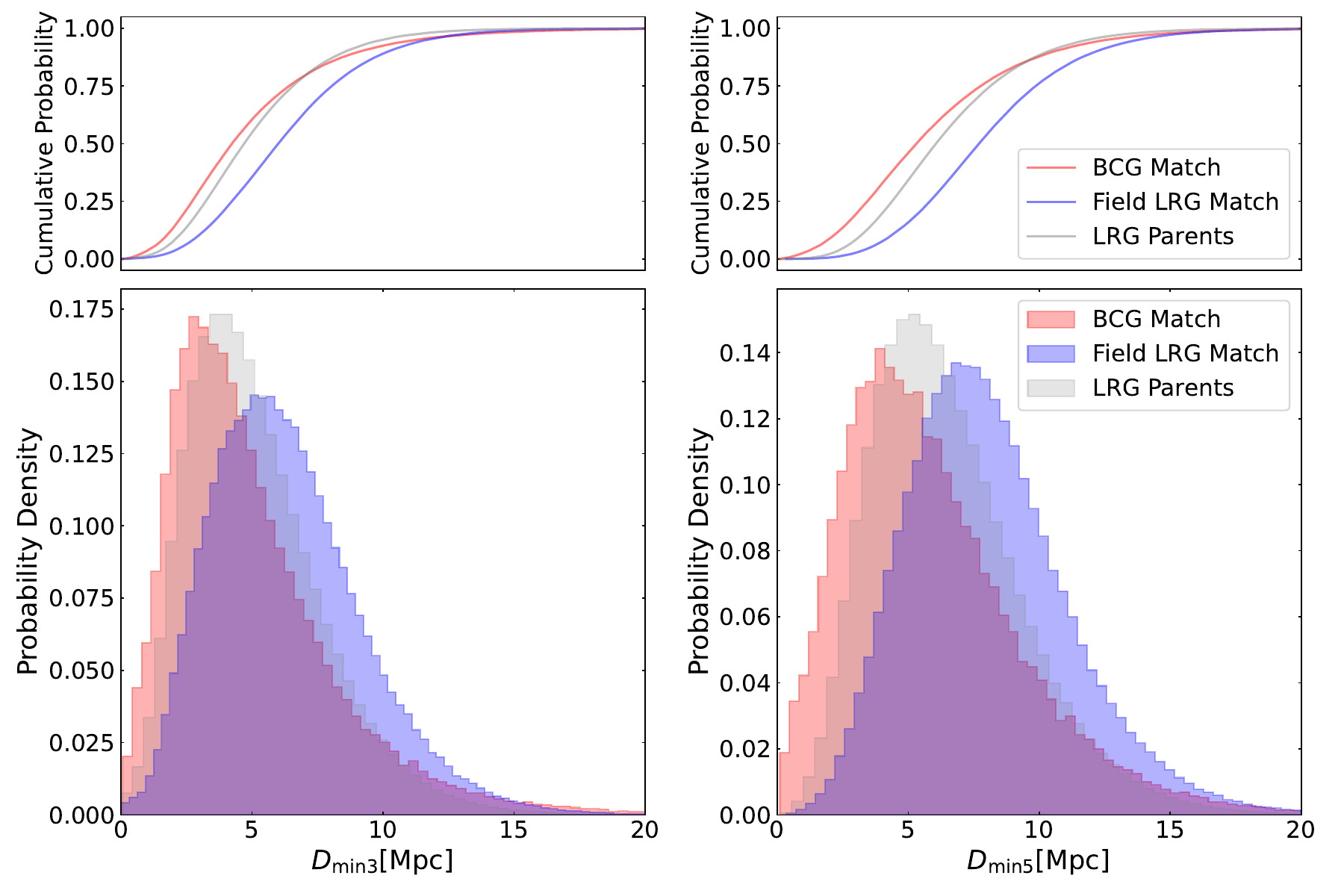}
\caption{Environmental density contrast between BCGs and LRGs. Cumulative distribution functions (top panels) and probability density functions (bottom panels) of the projected distances to the third- and fifth-nearest spectroscopic neighbours ($D_\mathrm{min3}$ and $D_\mathrm{min5}$) are shown on the left and right, respectively. BCGs (red) systematically reside in denser environments than matched field LRGs (blue) and the parent LRG sample (light grey).}
\label{fig:Dmin3&5_allz}
\end{figure}

\begin{figure*}[ht!]
\centering
\includegraphics[width=1.0\textwidth]{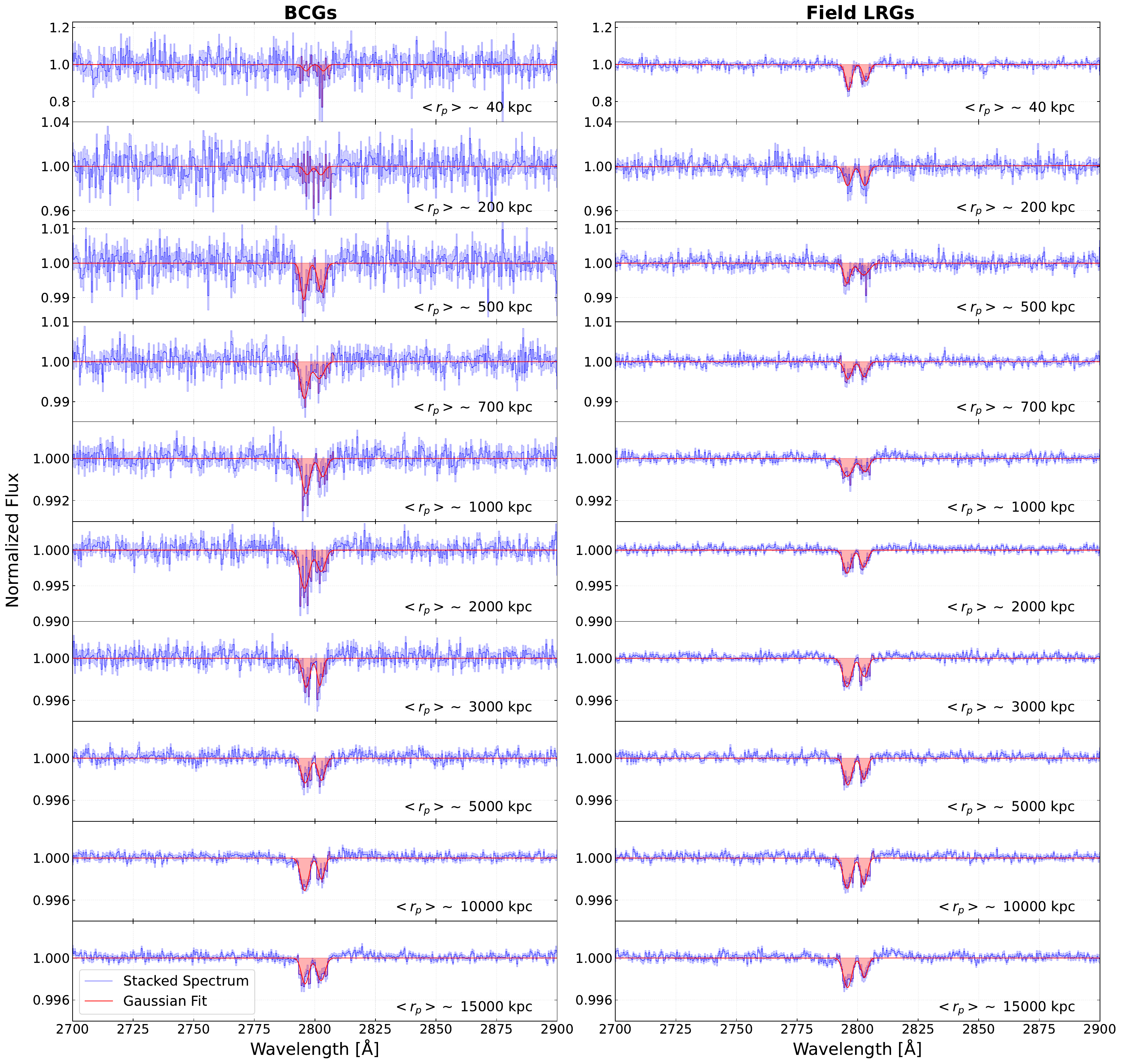}
\caption{Example of stacked background quasar spectra. The sequence of panels displays the stacked spectra across increasing projected distance $r_\mathrm{p}$ for BCGs (left) to those of field LRGs (right). The blue lines and shaded regions show the median spectra and standard deviations; the red curves are double-Gaussian fits, with the integration ranges for equivalent width measurements shaded in light red. \label{fig:EWs_example}}
\end{figure*}

\subsection{Environmental Density Quantification} \label{subsec:env}

The local galaxy density contrast between the BCG and field LRG samples is quantitatively verified using the projected comoving distance to the third- and fifth-neareast spectroscopic neighbors ($D_{\mathrm{min3}}$ and $D_{\mathrm{min5}}$) as density proxies (Figure \ref{fig:Dmin3&5_allz}). 
The cumulative distribution of BCGs is systematically shifted towards smaller separations compared to the field LRG sample, confirming they reside in denser environments. The probability density functions further reveal a pronounced peak for BCGs at 2–3 Mpc, while the distribution for field LRGs is broader and centred at larger distances. In contrast, the distribution for an unfiltered parent LRG sample lies between the two, indicating that a substantial fraction of typical LRGs reside in overdense environments, thereby validating our isolation criteria for defining a genuine field-galaxy control sample.

\section{Spectral Stacking Methodology} \label{sec:methods}
We measure the strength of Mg II absorption by computing the rest-frame equivalent width (EW) from stacked background quasar spectra. The galaxy samples are spatially cross-matched with background quasar sightlines, retaining all pairs with an impact parameter $r_\mathrm{p} < 15$ Mpc. This large matching radius allows us to trace gas from the inner CGM out to the cluster outskirts and general IGM, offering insight into environmental influences on gas over large scales. To avoid contamination from the quasar proximity region and possible broad absorption lines (BALs), we require each background quasar to be sufficiently separated from the foreground galaxy in redshift space with $\Delta z > 0.1$, which corresponds to a velocity separation of $\Delta v > 11000$ km/s. Moreover, because most of our sample lies at relatively low redshifts, the Mg II absorption features do not fall within the Ly$\alpha$ forest region of the background quasar.

Each quasar spectrum is continuum-fitted using the \texttt{QSOfit} package\footnote{\url{https://github.com/legolason/PyQSOFit}} \citep{QSOfit1, QSOfit2}, normalized by the fitted continuum, and smoothed with a median filter (71-pixel window) to reduce high-frequency noise and weak contaminating features \citep{2013ApJ...770..130Z, 2025ApJ...981...81C}. Spectra are shifted to the rest frame of the foreground galaxy, rebinned to a uniform {0.5 \AA} grid over 2700--2900 {\AA}, and stacked in impact-parameter bins by taking the median flux in each wavelength bin. Considering the possible contamination from the Lyα forest of background quasars, we find that only 0.02\% of our sample is affected, occurring exclusively at redshifts $z\gtrsim1.3$. Owing to our median stacking method, these rare contaminated sightlines are effectively treated as outliers and thus contribute negligibly to the final stacked spectra.

The Mg II absorption feature in each stacked spectrum is modeled with a physically constrained double-Gaussian profile, whose parameters are inferred via MCMC sampling using the \texttt{emcee} package \footnote{\url{https://emcee.readthedocs.io/en/stable/}} \citep{2013PASP..125..306F}. The two components are fixed at the known Mg II doublet separation and are required to share the same velocity offset and line width. We also impose physically motivated priors on the absorption amplitudes with EW(Mg II $\lambda$2796) $\geq$ EW(Mg II $\lambda$2803).

To assess whether the Mg II double-Gaussian model is statistically significant, we additionally compare the constrained double-Gaussian model with a zero-order polynomial continuum model using the Akaike information criteria (AIC, \citealt{1974ITAC...19..716A}). This test shows that the first seven BCG radial bins do not show strong statistical evidence favoring the Mg II double-Gaussian model. Therefore, we do not interpret these bins as individual significant detections. Nevertheless, the EW values reported in the radial profiles are retained as forced measurements at the expected Mg II doublet wavelengths, rather than as blind line detections. The Mg II doublet wavelengths are fixed a priori based on the foreground-galaxy redshift, and the fitted model is physically constrained: the two Mg II Gaussian components, sharing the same velocity offset and line width. Under this predefined generative model, the MCMC fitting provides a effective posterior estimate of the Mg II equivalent width even in low-S/N bins \citep{2010arXiv1008.4686H}. Similar forced-measurement strategies have been adopted in previous Mg II galaxy--QSO pair studies, where the EW is evaluated at the expected absorption wavelength without requiring a significant detection in each individual spectrum or pair \citep{Wu_2025ApJ...983..186W}.

Figure \ref{fig:EWs_example} shows the median composite spectra for BCGs (left) and matched field LRGs (right) in bins of $r_\mathrm{p}$. The stacking adopts an exponential binning scheme at small separations ($r_\mathrm{p} < 1.2$ Mpc) for comparison with earlier work \citep{Lan2018ApJ} and a shallower exponential binning at larger distances (out to 15 Mpc) to resolve radial trends in the outskirts. The spectra and their standard deviations are displayed together with the double-Gaussian fits. A clear environmental dependence is seen: within the inner region, Mg II absorption is weaker around BCGs than around field LRGs, whereas the trend reverses at larger radii. At the largest projected distances probed, the absorption strengths of the two samples converge to a comparable level.

\section{Results and Analyses} \label{sec:results}

\subsection{Environmental Dependence of Cool Gas} \label{subsec:environment}

We investigate how environmental influences cool gas distributions by comparing matched BCG and field LRG samples. The observed radial profiles of the total Mg II equivalent width can be described using the empirical ``two-halo" model introduced by \citet{Anand_2022}, which combines a power-law component with an exponential component:
\begin{equation}
    f(r_\mathrm{p}) = f_\mathrm{a} (r_\mathrm{p}/r_o)^{\alpha} + f_\mathrm{b} e^{-(r_\mathrm{p}/r_\mathrm{o})} + \beta. \label{eq:fitting}
\end{equation}
This model contains five free parameters: $f_\mathrm{a}$, $f_\mathrm{b}$, $\beta$, $r_\mathrm{o}$ and $\alpha$. The scale $r_\mathrm{o}$ marks the characteristic transition between the two regimes, with the slope governed by $\alpha$. We include an additional constant term $\beta$ to account for the asymptotic behavior of the equivalent width at large impact parameters. The best-fit parameters for the BCG and field LRG samples and corresponding subsamples, obtained via MCMC sampling, are presented in Table \ref{tab:fit_results}. The fitted model parameters may exhibit some degeneracy due to the limited number of data points and the associated measurement uncertainties. Nevertheless, the model curves provide a useful visualization that facilitates comparison of absorption profiles across different samples and helps to highlight the systematic differences in the observed trends.

\begin{figure*}[ht!]
\centering
\includegraphics[width=1.0\textwidth]{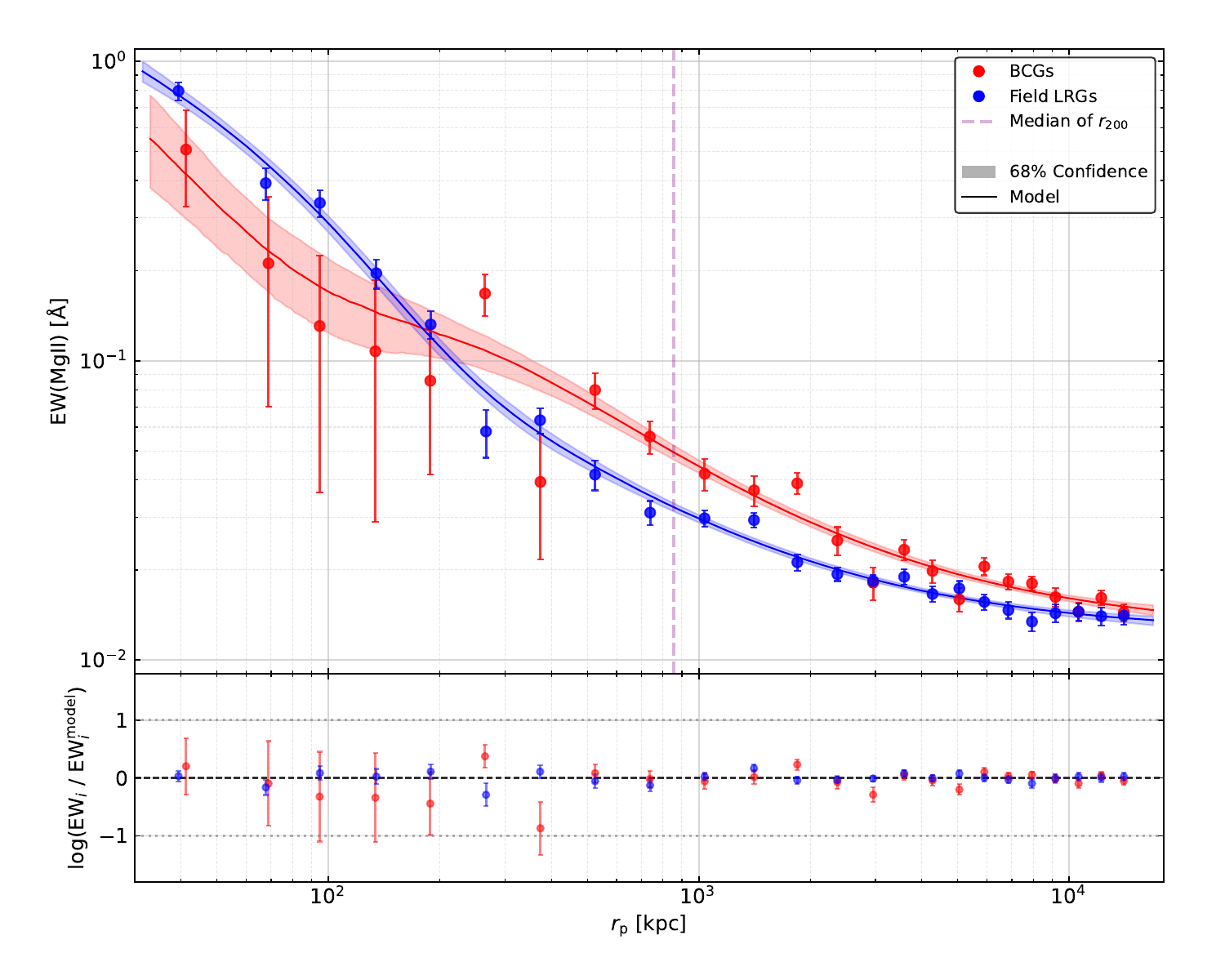}
\caption{Radial profiles of Mg II absorption around BCGs and field LRGs. Data points for BCGs (red) and field LRGs (blue) are plotted with 1$\sigma$ vertical error bars. The best-fit curves from the two‑halo model are overplotted for each sample, with the associated uncertainty represented by the shaded band. The median value of the virial radius $\tilde{r}_{200} \sim 855$ kpc of our sample is shown in purple dashed line. The bottom panels show the residuals between the measured equivalent widths and the model predictions. \label{fig:fit}}
\end{figure*}

\begin{table*}[ht]
\caption{Best-fit parameters of the two-halo model for BCGs and field LRGs}\label{tab:fit_results}%
\begin{tabular}{@{}lccccc@{}}
\toprule
\makechange{ } & $f_\mathrm{a}$ & $\alpha$ & $f_\mathrm{b}$ & $r_\mathrm{o}$ & $\beta$\\
\addlinespace[0.5em]
\hline \hline
\addlinespace[0.5em]
\textbf{BCGs (All Sample)} & $0.40^{+0.19}_{-0.14}$ & $-0.96^{+0.12}_{-0.12}$ & $-0.52^{+0.38}_{-0.50}$ & $75.5^{+36.3}_{-30.7}$ & $0.013^{+0.001}_{-0.001}$
\\
\addlinespace[0.5em]
\hline
\addlinespace[0.5em]
\textbf{Field LRGs (All Sample)} & $0.27^{+0.13}_{-0.08}$ & $-0.94^{+0.10}_{-0.10}$ & $0.83^{+0.22}_{-0.24}$ & $52.7^{+8.7}_{-8.9}$ & $0.012^{+0.001}_{-0.001}$
\\
\addlinespace[0.5em]
\hline
\addlinespace[0.5em]
\textbf{BCGs ($\mathrm{log}(M_{*}/M_{\odot}) < 11.3$)} & $1.75^{+0.77}_{-1.00}$ & $-1.37^{+0.17}_{-0.13}$ & $-2.44^{+2.23}_{-1.75}$ & $40.7^{+16.5}_{-7.4}$ & $0.014^{+0.001}_{-0.001}$
\\
\addlinespace[0.5em]
\hline
\addlinespace[0.5em]
\textbf{BCGs ($\mathrm{log}(M_{*}/M_{\odot}) \geq 11.3$)} & $0.47^{+0.24}_{-0.18}$ & $-0.93^{+0.14}_{-0.14}$ & $-0.80^{+0.56}_{-0.73}$ & $61.1^{+26.5}_{-20.6}$ & $0.012^{+0.001}_{-0.002}$
\\
\addlinespace[0.5em]
\hline
\addlinespace[0.5em]
\textbf{Field LRGs ($\mathrm{log}(M_{*}/M_{\odot}) < 11.3$)} & $0.52^{+0.30}_{-0.18}$ & $-1.19^{+0.14}_{-0.14}$ & $1.51^{+0.55}_{-0.67}$ & $44.8^{+8.5}_{-7.9}$ & $0.014^{+0.001}_{-0.002}$
\\
\addlinespace[0.5em]
\hline
\addlinespace[0.5em]
\textbf{Field LRGs ($\mathrm{log}(M_{*}/M_{\odot}) \geq 11.3$)}& $0.19^{+0.23}_{-0.07}$ & $-0.92^{+0.15}_{-0.13}$ & $0.35^{+0.21}_{-0.26}$ & $69.6^{+24.7}_{-26.7}$ & $0.012^{+0.001}_{-0.002}$
\\
\addlinespace[0.5em]
\botrule
\end{tabular}
\end{table*} 

The resulting radial profiles of the total Mg II equivalent width (EW; EW$_{2796}$ $+$ EW$_{2803}$) for both samples at a median redshift of $z\approx0.55$ are presented in Figure \ref{fig:fit}. Fitting the ``two-halo" model to the data reveals three distinct environmental regimes in the radial distribution of cool gas around the galaxies:
\begin{itemize}
    \item \textbf{Inner CGM ($r_\mathrm{p} \lesssim 200$ kpc):} Field LRGs show stronger Mg II absorption ($\sim 0.8$ \AA) than BCGs ($\sim 0.5$ \AA), corresponding to a $\sim 37.5$\% suppression around BCGs. This is consistent with earlier studies \citep{Anand_2021, Anand_2022} and likely results from efficient stripping and heating of cool gas by the hot ICM via ram pressure and thermal evaporation, as well as merger-driven heating \citep{2007MNRAS.382.1481N, 2022A&ARv..30....3B}. The decline in cool gas toward the cluster center aligns with hydrodynamical simulations that show rising ICM pressure and temperature in the inner regions \citep[e.g.,][]{2014Natur.509..177V, 2024MNRAS.527.1301R}.
    
    \item \textbf{Transition Region ($200$ kpc $\lesssim r_\mathrm{p} \lesssim$ 10 Mpc):} A notable reversal occurs at these intermediate distances, with BCGs exhibiting stronger Mg II absorption than field LRGs. The enhancement extends both within the halo and into its environs, signaling an increased cool-gas reservoir in the cluster outskirts. This finding agrees with simulations of large-scale filaments feeding clusters \citep{2024MNRAS.527.1301R}. Filaments can act as shielded channels, preserving cool, low-entropy gas during accretion onto the cluster halo.
 
    \item \textbf{Outer IGM ($r_\mathrm{p} \gtrsim$ 10 Mpc):} The absorption difference between the two samples diminishes, with both converging to a common baseline of about 0.014 \AA. 
    This convergence indicates that the cool, metal-enriched gas in the diffuse IGM is largely independent of the central galactic environment, representing a baseline level of chemical enrichment on large intergalactic scales.
    
\end{itemize} 
 
These results bridge galactic and cosmological scales, indicating that the local environment governs cool-gas distribution from the CGM to the cosmic web. The convergence of Mg II profiles in the outer IGM across different tracers validates the consistency of the analytical methodology, whereas the systematic differences in the inner and transition regions reflect fundamental distinctions in sample selection and intrinsic properties.

Our findings highlight two key features for BCGs relative to field LRGs: (1) a deficit of cool gas in the inner CGM, and (2) a significantly stronger, extended gas component in the outer halo to IGM. This points to a richer reservoir of cool gas in the large-scale cluster environment. These observations align with recent theoretical work. Simulations from TNG-Cluster and TNG300 reveal that cool gas in $z= 0$ clusters is ubiquitous, clumpy, and predominantly located in the outskirts, originating mainly from the recent accretion of pre-cooled gas via satellites \citep{2025arXiv250301960S}. Our detection of a strong, extended cool-gas component around BCGs supports this picture of an accretion-dominated origin, where cosmic-web filaments supply cool gas to the extended halo and potentially to the inner CGM of the central galaxy. However, the observed excess in the outer regions may also be partly contributed by intra‑halo clustering within massive halos; quantifying this effect will require complete spectroscopic observations of galaxies in and around clusters.

\subsection{Stellar mass dependence of Mg II absorption} \label{subsec:mass}

\begin{figure*}
\hspace{-1.7cm}
\includegraphics[width=1.2\textwidth]{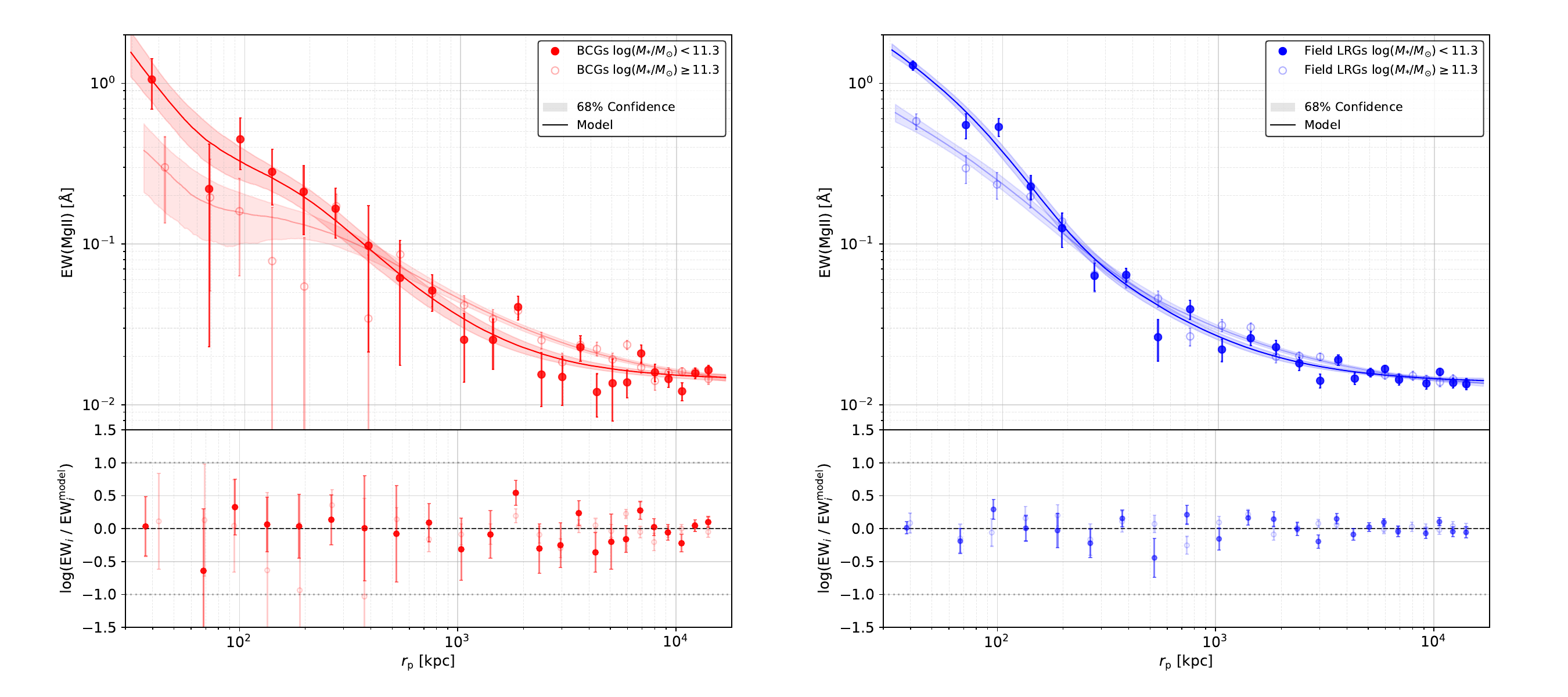}
\caption{Radial profiles of Mg II absorption for BCGs and field LRGs in low- and high-stellar-mass bins. Solid points indicate the low-mass bin ($\mathrm{log}(M_{*}/M_{\odot}) < 11.3$), while hollow points represent the high-mass bin ($\mathrm{log}(M_{*}/M_{\odot}) \geq 11.3$). This figure follows the same plotting conventions as Figure \ref{fig:fit}.
\label{fig:m}}
\end{figure*}

To further probe the drivers of cool gas distribution, we investigate the relationship between Mg II absorption and stellar mass by dividing both the BCG and field LRG samples at the median stellar mass of $\mathrm{log}(M_{*}/M_{\odot}) = 11.3$ into low-mass and high-mass subsets. The radial profiles of Mg II equivalent width for these mass bins are shown in Fig. \ref{fig:m}. The data reveal how the radial distribution of cool gas is modulated by both stellar mass and environment across three regimes.

In the inner CGM ($r_\mathrm{p} \lesssim 200$ kpc), low-mass galaxies exhibit systematically stronger Mg II absorption than their high-mass counterparts, a trend seen in both BCGs (albeit with larger uncertainty) and field LRGs. In this region, field LRGs also show stronger absorption than BCGs of comparable mass, reinforcing the environmental suppression described earlier.

Within the transition region, the environmental signal dominates: BCGs maintain a higher cool-gas abundance than field LRGs overall. Moreover, a tentative trend suggests that more massive BCGs are associated with enhanced cool-gas content here, despite the substantial measurement uncertainties. This hint aligns with simulations indicating that cluster connectivity, the number of filaments linked to a cluster, correlates strongly with total cluster mass \citep{2024A&A...692A..44S}. More massive BCGs, residing in better-connected clusters, may therefore have access to larger reservoirs of cool gas accreted along filaments.

At sufficiently large scales, differences linked to galaxy type and stellar mass disappear. All radial profiles converge to a common Mg II equivalent width of $\sim 0.014$ \AA. This convergence indicates that the Mg II absorption on these scales is independent of the properties of the central galaxy and its immediate environment.

\section{Summary} \label{sec:summary}

In summary, we present the first statistical mapping of cool gas around massive quiescent galaxies at $z\approx0.55$ using Mg II absorption from stacking sight-line spectra of over $10^6$ DESI background quasars. A direct comparison between BCGs and matched field LRGs reveals how the cluster environment regulates gaseous halos from $\sim 40$ kpc out to the outer IGM ($\sim$15 Mpc). The most striking result is the enhanced cool gas reservoir detected around BCGs at 200 kpc -- 10 Mpc, which challenges the simplified view of clusters as purely destructive environments. This indicates that clusters might play a dual role: they efficiently deplete the gas in the inner CGM while simultaneously assembling cool gas at larger scales. This work provides new observational constraints for models of baryon cycling and galaxy quenching, highlighting the possible role of cosmic-web feeding in shaping the circumgalactic halos of the most massive galaxies in the universe. Furthermore, by splitting the sample by stellar mass, we show how gas distribution is modulated by both mass and environment across different radial regimes.

\begin{acknowledgements}

We acknowledge Cheng Li and Furen Deng for helpful discussions. This work is supported by the National Key R\&D Program of China (Grant Nos. 2023YFA1607804), the National Natural Science Foundation of China (NSFC; grant Nos. 12120101003, 12373010, 12473008, 12233008, and 12503019), and China Manned Space Project (No. CMS-CSST-2025-A06). The authors also acknowledge the supports from the National Key R\&D Program of China (Grant Nos. 2022YFA1602902, 2023YFA1607800, 2023YFA1608100, and 2025YFE0202300), the Strategic Priority Research Program of the Chinese Academy of Sciences (Grant Nos. XDB0550100 and XDB0550000) and the Programs of National Astronomical Observatories Chinese Academy of Sciences (Grant Nos. E5ZQ7801, E5ZB7801, and E4TG2001). 

RY and TF are supported by the National SKA Program of China No. 2025SKA0150103, National Natural Science Foundation of China under Nos. 12550002, 12133008, 12221003, 11890692. We acknowledge the science research grants from the China Manned Space Project with No. CMS-CSST-2021-A04 and No. CMS-CSST-2025-A10.

This research used data obtained with the Dark Energy Spectroscopic Instrument (DESI). DESI construction and operations is managed by the Lawrence Berkeley National Laboratory. This material is based upon work supported by the U.S. Department of Energy, Office of Science, Office of High-Energy Physics, under Contract No. DE–AC02–05CH11231, and by the National Energy Research Scientific Computing Center, a DOE Office of Science User Facility under the same contract. Additional support for DESI was provided by the U.S. National Science Foundation (NSF), Division of Astronomical Sciences under Contract No. AST-0950945 to the NSF’s National Optical-Infrared Astronomy Research Laboratory; the Science and Technology Facilities Council of the United Kingdom; the Gordon and Betty Moore Foundation; the Heising-Simons Foundation; the French Alternative Energies and Atomic Energy Commission (CEA); the National Council of Humanities, Science and Technology of Mexico (CONAHCYT); the Ministry of Science and Innovation of Spain (MICINN), and by the DESI Member Institutions: \url{www.desi.lbl.gov/collaborating-institutions}. The DESI collaboration is honored to be permitted to conduct scientific research on I’oligam Du’ag (Kitt Peak), a mountain with particular significance to the Tohono O’odham Nation. Any opinions, findings, and conclusions or recommendations expressed in this material are those of the author (s) and do not necessarily reflect the views of the U.S. National Science Foundation, the U.S. Department of Energy, or any of the listed funding agencies.

\end{acknowledgements}

\vspace{3em}
\appendix

\section{Control Tests for the LRG Sample Size} \label{sec: size}

\renewcommand{\thefigure}{\thesection \arabic{figure}}
\setcounter{figure}{0}

\begin{figure}[ht!]
\centering
\includegraphics[width=1.0\textwidth]{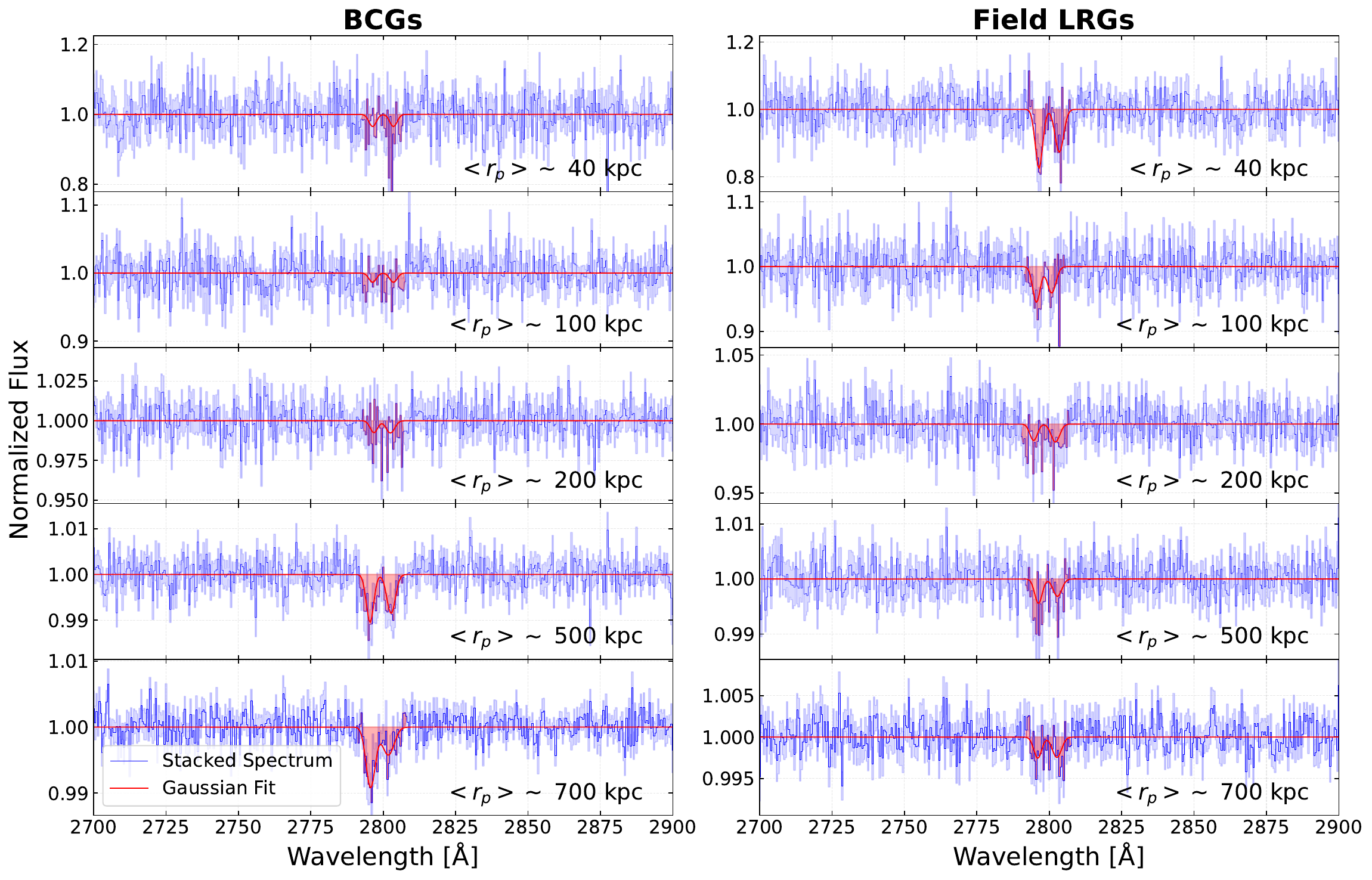}
\caption{Stacked background quasar spectra for the BCG and matched field LRG sample with equalized sizes. This figure is analogous to Figure \ref{fig:EWs_example} but shows the field LRG sample after one‑to‑one matching in stellar mass and redshift with the BCG sample. Both samples now contain an equal number of sources (72,913). All plotting conventions (line styles and colors) follow Figure \ref{fig:EWs_example}. \label{fig:EWs_example_eq}}
\end{figure}

\begin{figure*}[ht!]
\centering
\includegraphics[width=1.0\textwidth]{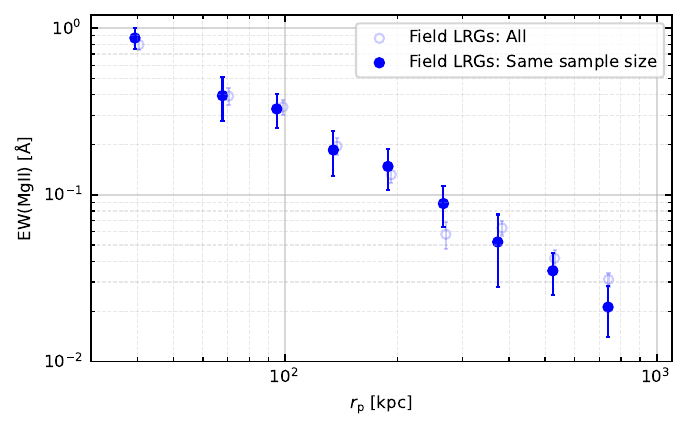}
\caption{Radial profile of Mg II absorption for two control samples of field LRGs. The originally selected LRGs and a subset matched one‑to‑one in stellar mass and redshift to the BCG sample are displayed in light blue and blue, respectively. \label{fig:eqEWs_distri}}
\end{figure*}

To assess whether the larger sample size of field LRGs (approximately three times that of BCGs; see Section \ref{subsec:fields}) biases our comparison by enhancing their stacked signal-to-noise ratio (S/N), we constructed a control sample through one-to-one matching in both stellar mass and redshift. This yielded two samples of equal size ($N=72,913$ each). The median-stacked spectra around the Mg II absorption lines for these matched samples are presented in Figure \ref{fig:EWs_example_eq}. With matched sample sizes, the stacked spectra of field LRGs and BCGs achieve comparable S/N. Crucially, the radial profile of Mg II equivalent width derived from the matched field LRG sample shows no significant deviation from that of the full field LRG sample (Figure \ref{fig:eqEWs_distri}). This consistency confirms that our primary results, particularly the suppressed Mg II absorption around BCGs in the inner CGM, are robust and are not driven by differences in sample size or stacking S/N.

\section{Choice of Distance Metric} \label{sec:norm}

\renewcommand{\thefigure}{\thesection \arabic{figure}}
\setcounter{figure}{0}

In this work, we present radial profiles using the projected distance $r_p$, rather than normalizing it by a galaxy-specific virial radius such as $r_{200}$. This choice is made for several reasons. First, robust estimates of $r_{200}$ for both BCGs and field LRGs carry substantial uncertainty. Normalizing by an imprecise $r_{200}$ would artificially smooth the intrinsic radial trends we aim to measure. Second, our sample spans a relatively narrow redshift range, where the benefits of such scaling are minimal compared to the added complexity. Finally, using $r_p$ directly maintains consistency with key previous studies of cool CGM gas \citep[e.g.,][]{2014MNRAS.439.3139Z, Lan2018ApJ, 2025ApJ...981...81C}, enabling a straightforward comparison of our results with established work.

\begin{figure}[ht!]
\centering
\includegraphics[width=1.0\textwidth]{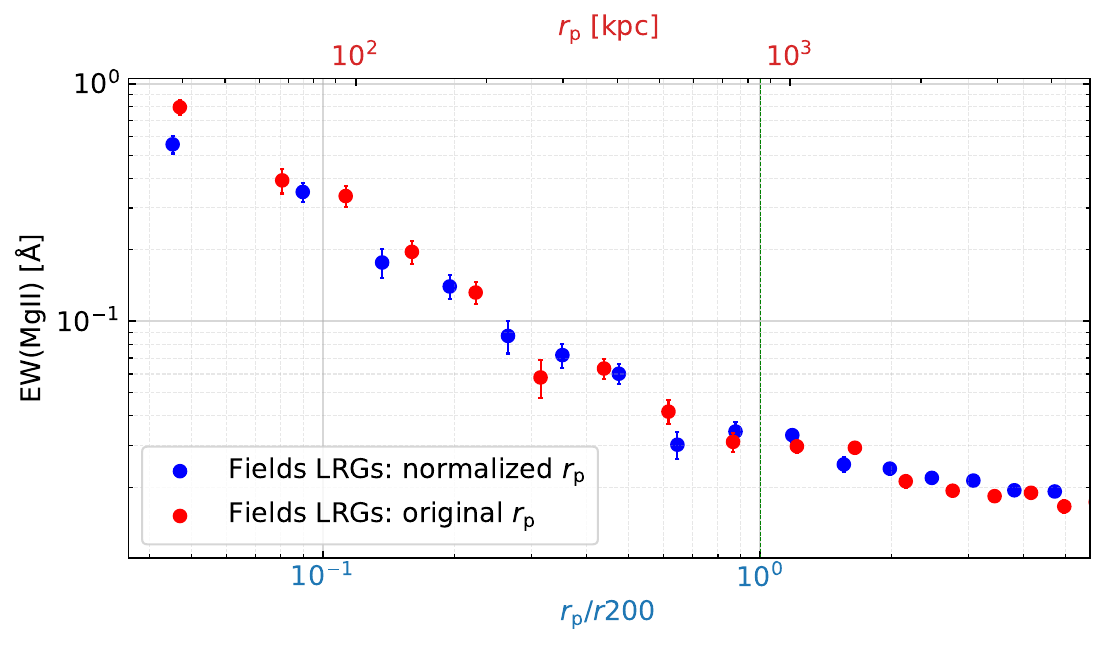}
\caption{Radial Mg II profile for field LRGs using both projected and virial-normalized distance scales. The equivalent width measurements (red circles) are plotted against projected distance $r_\mathrm{p}$ (top axis) and, for the same data, against normalized distance $r_\mathrm{p}/r_{200}$ (bottom axis). The vertical green dotted line marks the median virial radius $\tilde{r}_{200} \sim 855$ kpc.}. \label{fig:norm_or_not}
\end{figure}

To explicitly test the impact of this choice, we performed a comparative analysis. We estimate $r_{200}$ for field LRGs using the stellar-to-halo mass relation from \citet{SHMR} and the standard virial relation \citep{1998ApJ...495...80B, Anand_2021}:
\begin{equation}
    r_{200} = 211.83\ M_{12}^{1/3}[\Omega_\mathrm{m,0}(1+z)^{3} + \Omega_\mathrm{\Lambda, 0}]^{-1/3}\ [\mathrm{kpc}] \label{eq:r200}
\end{equation}
where $M_{12}$ is the halo mass in units of $10^{12} M_\odot$, $z$ is the redshift of the galaxy, and $\Omega_\mathrm{m,0}$ and $\Omega_\mathrm{\Lambda, 0}$ denote the present-day matter and dark-energy density parameters, respectively. The radial profile of Mg II equivalent width constructed using $r_p/r_{200}$ is then compared to the profile using $r_p$ directly. As shown in Figure \ref{fig:norm_or_not}, the two profiles show no significant difference in shape or trends. The normalization does not reveal new features and can introduce a slight smoothing effect due to the uncertainties in $r_{200}$. This test confirms that our core results are robust and insensitive to the distance metric, validating our use of the simple projected distance.

\section{Comparison with Previous SDSS LRG Profiles} \label{sec:compare}

\renewcommand{\thefigure}{\thesection \arabic{figure}}
\setcounter{figure}{0}

\begin{figure}[ht!]
\centering
\includegraphics[width=1.0\textwidth]{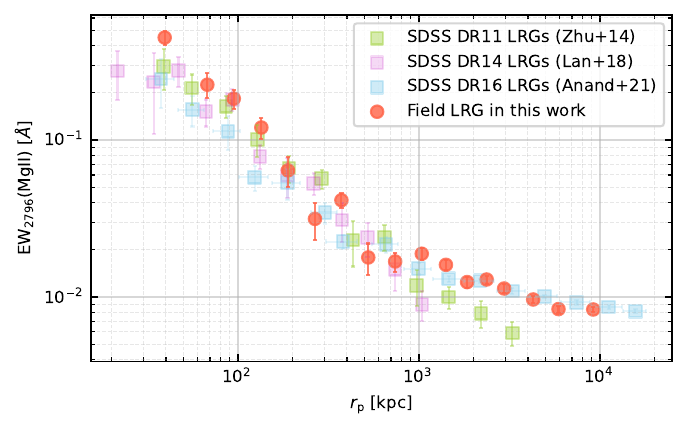}
\caption{Comparison of the Mg II $\mathrm{EW}_{2796}$ radial profile for LRGs from this work and from earlier SDSS studies. Measurements from SDSS DR11 (green squares; \citep{2014MNRAS.439.3139Z}), DR14 (magenta squares; \citep{Lan2018ApJ}), and DR16 (blue squares; \citep{Anand_2021}) are shown together with the results for the field LRG sample in this work (orange points). \label{fig:compare}}
\end{figure}

The radial profile of Mg II $\lambda\lambda 2796$ equivalent width for our field LRG sample is compared with earlier SDSS-based measurements in Figure \ref{fig:compare}. In the inner CGM ($r_\mathrm{p} \leq 200$ kpc), our field LRGs exhibit slightly enhanced Mg II absorption compared to the SDSS LRG samples. This marginal enhancement likely reflects an environmental difference: our sample is strictly selected to avoid clusters, meaning these LRGs reside in lower-density environments where galaxy–galaxy interactions are less frequent and less cool gas has been stripped from their halos. At larger projected distances, our measurements are in good agreement with those of LRG samples from SDSS DR16 \citep{Anand_2021} but remain systematically higher than those reported in other SDSS-based studies. These discrepancies likely stem from differences in sample selection, absorption-line measurement methodology, and stacking techniques across different works.

\bibliography{coolgas_cluster}{}

@ARTICLE{1986A&A...169....1B,
       author = {{Bergeron}, J. and {Stasi{\'n}ska}, G.},
        title = "{Absorption line systems in QSO spectra : properties derived from observations and from photoionization models.}",
      journal = {\aap},
         year = 1986,
        month = nov,
       volume = {169},
        pages = {1-3},
       adsurl = {https://ui.adsabs.harvard.edu/abs/1986A&A...169....1B}
}

@ARTICLE{Charlton_2003ApJ...589..111C,
       author = {{Charlton}, Jane C. and {Ding}, Jie and {Zonak}, Stephanie G. and {Churchill}, Christopher W. and {Bond}, Nicholas A. and {Rigby}, Jane R.},
        title = "{High-Resolution STIS/Hubble Space Telescope and HIRES/Keck Spectra of Three Weak Mg II Absorbers toward PG 1634+706}",
      journal = {\apj},
         year = 2003,
        month = may,
       volume = {589},
       number = {1},
        pages = {111-125},
          doi = {10.1086/374353},
archivePrefix = {arXiv},
       eprint = {astro-ph/0302394},
 primaryClass = {astro-ph},
       adsurl = {https://ui.adsabs.harvard.edu/abs/2003ApJ...589..111C}
}

@ARTICLE{EDR,
       author = {{DESI Collaboration} and {Adame}, A.~G. and {Aguilar}, J. and {Ahlen}, S. and {Alam}, S. and {Aldering}, G. and {Alexander}, D.~M. and {Alfarsy}, R. and {Allende Prieto}, C. and {Alvarez}, M. and {Alves}, O. and {Anand}, A. and {Andrade-Oliveira}, F. and {Armengaud}, E. and {Asorey}, J. and {Avila}, S. and {Aviles}, A. and {Bailey}, S. and {Balaguera-Antol{\'\i}nez}, A. and {Ballester}, O. and {Baltay}, C. and {Bault}, A. and {Bautista}, J. and {Behera}, J. and {Beltran}, S.~F. and {BenZvi}, S. and {Beraldo e Silva}, L. and {Bermejo-Climent}, J.~R. and {Berti}, A. and {Besuner}, R. and {Beutler}, F. and {Bianchi}, D. and {Blake}, C. and {Blum}, R. and {Bolton}, A.~S. and {Brieden}, S. and {Brodzeller}, A. and {Brooks}, D. and {Brown}, Z. and {Buckley-Geer}, E. and {Burtin}, E. and {Cabayol-Garcia}, L. and {Cai}, Z. and {Canning}, R. and {Cardiel-Sas}, L. and {Carnero Rosell}, A. and {Castander}, F.~J. and {Cervantes-Cota}, J.~L. and {Chabanier}, S. and {Chaussidon}, E. and {Chaves-Montero}, J. and {Chen}, S. and {Chen}, X. and {Chuang}, C. and {Claybaugh}, T. and {Cole}, S. and {Cooper}, A.~P. and {Cuceu}, A. and {Davis}, T.~M. and {Dawson}, K. and {de Belsunce}, R. and {de la Cruz}, R. and {de la Macorra}, A. and {Della Costa}, J. and {de Mattia}, A. and {Demina}, R. and {Demirbozan}, U. and {DeRose}, J. and {Dey}, A. and {Dey}, B. and {Dhungana}, G. and {Ding}, J. and {Ding}, Z. and {Doel}, P. and {Doshi}, R. and {Douglass}, K. and {Edge}, A. and {Eftekharzadeh}, S. and {Eisenstein}, D.~J. and {Elliott}, A. and {Ereza}, J. and {Escoffier}, S. and {Fagrelius}, P. and {Fan}, X. and {Fanning}, K. and {Fawcett}, V.~A. and {Ferraro}, S. and {Flaugher}, B. and {Font-Ribera}, A. and {Forero-Romero}, J.~E. and {Forero-S{\'a}nchez}, D. and {Frenk}, C.~S. and {G{\"a}nsicke}, B.~T. and {Garc{\'\i}a}, L. {\'A}. and {Garc{\'\i}a-Bellido}, J. and {Garcia-Quintero}, C. and {Garrison}, L.~H. and {Gil-Mar{\'\i}n}, H. and {Golden-Marx}, J. and {Gontcho A Gontcho}, S. and {Gonzalez-Morales}, A.~X. and {Gonzalez-Perez}, V. and {Gordon}, C. and {Graur}, O. and {Green}, D. and {Gruen}, D. and {Guy}, J. and {Hadzhiyska}, B. and {Hahn}, C. and {Han}, J.~J. and {Hanif}, M.~M.~S. and {Herrera-Alcantar}, H.~K. and {Honscheid}, K. and {Hou}, J. and {Howlett}, C. and {Huterer}, D. and {Ir{\v{s}}i{\v{c}}}, V. and {Ishak}, M. and {Jacques}, A. and {Jana}, A. and {Jiang}, L. and {Jimenez}, J. and {Jing}, Y.~P. and {Joudaki}, S. and {Joyce}, R. and {Jullo}, E. and {Juneau}, S. and {Kara{\c{c}}ayl{\i}}, N.~G. and {Karim}, T. and {Kehoe}, R. and {Kent}, S. and {Khederlarian}, A. and {Kim}, S. and {Kirkby}, D. and {Kisner}, T. and {Kitaura}, F. and {Kizhuprakkat}, N. and {Kneib}, J. and {Koposov}, S.~E. and {Kov{\'a}cs}, A. and {Kremin}, A. and {Krolewski}, A. and {L'Huillier}, B. and {Lahav}, O. and {Lambert}, A. and {Lamman}, C. and {Lan}, T. -W. and {Landriau}, M. and {Lang}, D. and {Lange}, J.~U. and {Lasker}, J. and {Leauthaud}, A. and {Le Guillou}, L. and {Levi}, M.~E. and {Li}, T.~S. and {Linder}, E. and {Lyons}, A. and {Magneville}, C. and {Manera}, M. and {Manser}, C.~J. and {Margala}, D. and {Martini}, P. and {McDonald}, P. and {Medina}, G.~E. and {Medina-Varela}, L. and {Meisner}, A. and {Mena-Fern{\'a}ndez}, J. and {Meneses-Rizo}, J. and {Mezcua}, M. and {Miquel}, R. and {Montero-Camacho}, P. and {Moon}, J. and {Moore}, S. and {Moustakas}, J. and {Mueller}, E. and {Mundet}, J. and {Mu{\~n}oz-Guti{\'e}rrez}, A. and {Myers}, A.~D. and {Nadathur}, S. and {Napolitano}, L. and {Neveux}, R. and {Newman}, J.~A. and {Nie}, J. and {Nikutta}, R. and {Niz}, G. and {Norberg}, P. and {Noriega}, H.~E. and {Paillas}, E. and {Palanque-Delabrouille}, N. and {Palmese}, A. and {Pan}, Z. and {Parkinson}, D. and {Penmetsa}, S. and {Percival}, W.~J. and {P{\'e}rez-Fern{\'a}ndez}, A. and {P{\'e}rez-R{\`a}fols}, I. and {Pieri}, M. and {Poppett}, C. and {Porredon}, A. and {Pothier}, S.},
        title = "{The Early Data Release of the Dark Energy Spectroscopic Instrument}",
      journal = {\aj},
         year = 2024,
        month = aug,
       volume = {168},
       number = {2},
          eid = {58},
        pages = {58},
          doi = {10.3847/1538-3881/ad3217},
archivePrefix = {arXiv},
       eprint = {2306.06308},
 primaryClass = {astro-ph.CO},
       adsurl = {https://ui.adsabs.harvard.edu/abs/2024AJ....168...58D}
}

@ARTICLE{PART1,
       author = {{DESI Collaboration} and {Aghamousa}, Amir and {Aguilar}, Jessica and {Ahlen}, Steve and {Alam}, Shadab and {Allen}, Lori E. and {Allende Prieto}, Carlos and {Annis}, James and {Bailey}, Stephen and {Balland}, Christophe and {Ballester}, Otger and {Baltay}, Charles and {Beaufore}, Lucas and {Bebek}, Chris and {Beers}, Timothy C. and {Bell}, Eric F. and {Bernal}, Jos{\'e} Luis and {Besuner}, Robert and {Beutler}, Florian and {Blake}, Chris and {Bleuler}, Hannes and {Blomqvist}, Michael and {Blum}, Robert and {Bolton}, Adam S. and {Briceno}, Cesar and {Brooks}, David and {Brownstein}, Joel R. and {Buckley-Geer}, Elizabeth and {Burden}, Angela and {Burtin}, Etienne and {Busca}, Nicolas G. and {Cahn}, Robert N. and {Cai}, Yan-Chuan and {Cardiel-Sas}, Laia and {Carlberg}, Raymond G. and {Carton}, Pierre-Henri and {Casas}, Ricard and {Castander}, Francisco J. and {Cervantes-Cota}, Jorge L. and {Claybaugh}, Todd M. and {Close}, Madeline and {Coker}, Carl T. and {Cole}, Shaun and {Comparat}, Johan and {Cooper}, Andrew P. and {Cousinou}, M. -C. and {Crocce}, Martin and {Cuby}, Jean-Gabriel and {Cunningham}, Daniel P. and {Davis}, Tamara M. and {Dawson}, Kyle S. and {de la Macorra}, Axel and {De Vicente}, Juan and {Delubac}, Timoth{\'e}e and {Derwent}, Mark and {Dey}, Arjun and {Dhungana}, Govinda and {Ding}, Zhejie and {Doel}, Peter and {Duan}, Yutong T. and {Ealet}, Anne and {Edelstein}, Jerry and {Eftekharzadeh}, Sarah and {Eisenstein}, Daniel J. and {Elliott}, Ann and {Escoffier}, St{\'e}phanie and {Evatt}, Matthew and {Fagrelius}, Parker and {Fan}, Xiaohui and {Fanning}, Kevin and {Farahi}, Arya and {Farihi}, Jay and {Favole}, Ginevra and {Feng}, Yu and {Fernandez}, Enrique and {Findlay}, Joseph R. and {Finkbeiner}, Douglas P. and {Fitzpatrick}, Michael J. and {Flaugher}, Brenna and {Flender}, Samuel and {Font-Ribera}, Andreu and {Forero-Romero}, Jaime E. and {Fosalba}, Pablo and {Frenk}, Carlos S. and {Fumagalli}, Michele and {Gaensicke}, Boris T. and {Gallo}, Giuseppe and {Garcia-Bellido}, Juan and {Gaztanaga}, Enrique and {Pietro Gentile Fusillo}, Nicola and {Gerard}, Terry and {Gershkovich}, Irena and {Giannantonio}, Tommaso and {Gillet}, Denis and {Gonzalez-de-Rivera}, Guillermo and {Gonzalez-Perez}, Violeta and {Gott}, Shelby and {Graur}, Or and {Gutierrez}, Gaston and {Guy}, Julien and {Habib}, Salman and {Heetderks}, Henry and {Heetderks}, Ian and {Heitmann}, Katrin and {Hellwing}, Wojciech A. and {Herrera}, David A. and {Ho}, Shirley and {Holland}, Stephen and {Honscheid}, Klaus and {Huff}, Eric and {Hutchinson}, Timothy A. and {Huterer}, Dragan and {Hwang}, Ho Seong and {Illa Laguna}, Joseph Maria and {Ishikawa}, Yuzo and {Jacobs}, Dianna and {Jeffrey}, Niall and {Jelinsky}, Patrick and {Jennings}, Elise and {Jiang}, Linhua and {Jimenez}, Jorge and {Johnson}, Jennifer and {Joyce}, Richard and {Jullo}, Eric and {Juneau}, St{\'e}phanie and {Kama}, Sami and {Karcher}, Armin and {Karkar}, Sonia and {Kehoe}, Robert and {Kennamer}, Noble and {Kent}, Stephen and {Kilbinger}, Martin and {Kim}, Alex G. and {Kirkby}, David and {Kisner}, Theodore and {Kitanidis}, Ellie and {Kneib}, Jean-Paul and {Koposov}, Sergey and {Kovacs}, Eve and {Koyama}, Kazuya and {Kremin}, Anthony and {Kron}, Richard and {Kronig}, Luzius and {Kueter-Young}, Andrea and {Lacey}, Cedric G. and {Lafever}, Robin and {Lahav}, Ofer and {Lambert}, Andrew and {Lampton}, Michael and {Landriau}, Martin and {Lang}, Dustin and {Lauer}, Tod R. and {Le Goff}, Jean-Marc and {Le Guillou}, Laurent and {Le Van Suu}, Auguste and {Lee}, Jae Hyeon and {Lee}, Su-Jeong and {Leitner}, Daniela and {Lesser}, Michael and {Levi}, Michael E. and {L'Huillier}, Benjamin and {Li}, Baojiu and {Liang}, Ming and {Lin}, Huan and {Linder}, Eric and {Loebman}, Sarah R. and {Luki{\'c}}, Zarija and {Ma}, Jun and {MacCrann}, Niall and {Magneville}, Christophe and {Makarem}, Laleh and {Manera}, Marc and {Manser}, Christopher J. and {Marshall}, Robert and {Martini}, Paul and {Massey}, Richard and {Matheson}, Thomas and {McCauley}, Jeremy and {McDonald}, Patrick and {McGreer}, Ian D. and {Meisner}, Aaron and {Metcalfe}, Nigel and {Miller}, Timothy N. and {Miquel}, Ramon and {Moustakas}, John and {Myers}, Adam and {Naik}, Milind and {Newman}, Jeffrey A. and {Nichol}, Robert C. and {Nicola}, Andrina and {Nicolati da Costa}, Luiz and {Nie}, Jundan and {Niz}, Gustavo and {Norberg}, Peder and {Nord}, Brian and {Norman}, Dara and {Nugent}, Peter and {O'Brien}, Thomas and {Oh}, Minji and {Olsen}, Knut A.~G.},
        title = "{The DESI Experiment Part I: Science,Targeting, and Survey Design}",
      journal = {arXiv e-prints},
         year = 2016,
        month = oct,
          eid = {arXiv:1611.00036},
        pages = {arXiv:1611.00036},
          doi = {10.48550/arXiv.1611.00036},
archivePrefix = {arXiv},
       eprint = {1611.00036},
 primaryClass = {astro-ph.IM},
       adsurl = {https://ui.adsabs.harvard.edu/abs/2016arXiv161100036D}
}

@ARTICLE{Planck18,
       author = {{Planck Collaboration} and {Aghanim}, N. and {Akrami}, Y. and {Ashdown}, M. and {Aumont}, J. and {Baccigalupi}, C. and {Ballardini}, M. and {Banday}, A.~J. and {Barreiro}, R.~B. and {Bartolo}, N. and {Basak}, S. and {Battye}, R. and {Benabed}, K. and {Bernard}, J. -P. and {Bersanelli}, M. and {Bielewicz}, P. and {Bock}, J.~J. and {Bond}, J.~R. and {Borrill}, J. and {Bouchet}, F.~R. and {Boulanger}, F. and {Bucher}, M. and {Burigana}, C. and {Butler}, R.~C. and {Calabrese}, E. and {Cardoso}, J. -F. and {Carron}, J. and {Challinor}, A. and {Chiang}, H.~C. and {Chluba}, J. and {Colombo}, L.~P.~L. and {Combet}, C. and {Contreras}, D. and {Crill}, B.~P. and {Cuttaia}, F. and {de Bernardis}, P. and {de Zotti}, G. and {Delabrouille}, J. and {Delouis}, J. -M. and {Di Valentino}, E. and {Diego}, J.~M. and {Dor{\'e}}, O. and {Douspis}, M. and {Ducout}, A. and {Dupac}, X. and {Dusini}, S. and {Efstathiou}, G. and {Elsner}, F. and {En{\ss}lin}, T.~A. and {Eriksen}, H.~K. and {Fantaye}, Y. and {Farhang}, M. and {Fergusson}, J. and {Fernandez-Cobos}, R. and {Finelli}, F. and {Forastieri}, F. and {Frailis}, M. and {Fraisse}, A.~A. and {Franceschi}, E. and {Frolov}, A. and {Galeotta}, S. and {Galli}, S. and {Ganga}, K. and {G{\'e}nova-Santos}, R.~T. and {Gerbino}, M. and {Ghosh}, T. and {Gonz{\'a}lez-Nuevo}, J. and {G{\'o}rski}, K.~M. and {Gratton}, S. and {Gruppuso}, A. and {Gudmundsson}, J.~E. and {Hamann}, J. and {Handley}, W. and {Hansen}, F.~K. and {Herranz}, D. and {Hildebrandt}, S.~R. and {Hivon}, E. and {Huang}, Z. and {Jaffe}, A.~H. and {Jones}, W.~C. and {Karakci}, A. and {Keih{\"a}nen}, E. and {Keskitalo}, R. and {Kiiveri}, K. and {Kim}, J. and {Kisner}, T.~S. and {Knox}, L. and {Krachmalnicoff}, N. and {Kunz}, M. and {Kurki-Suonio}, H. and {Lagache}, G. and {Lamarre}, J. -M. and {Lasenby}, A. and {Lattanzi}, M. and {Lawrence}, C.~R. and {Le Jeune}, M. and {Lemos}, P. and {Lesgourgues}, J. and {Levrier}, F. and {Lewis}, A. and {Liguori}, M. and {Lilje}, P.~B. and {Lilley}, M. and {Lindholm}, V. and {L{\'o}pez-Caniego}, M. and {Lubin}, P.~M. and {Ma}, Y. -Z. and {Mac{\'\i}as-P{\'e}rez}, J.~F. and {Maggio}, G. and {Maino}, D. and {Mandolesi}, N. and {Mangilli}, A. and {Marcos-Caballero}, A. and {Maris}, M. and {Martin}, P.~G. and {Martinelli}, M. and {Mart{\'\i}nez-Gonz{\'a}lez}, E. and {Matarrese}, S. and {Mauri}, N. and {McEwen}, J.~D. and {Meinhold}, P.~R. and {Melchiorri}, A. and {Mennella}, A. and {Migliaccio}, M. and {Millea}, M. and {Mitra}, S. and {Miville-Desch{\^e}nes}, M. -A. and {Molinari}, D. and {Montier}, L. and {Morgante}, G. and {Moss}, A. and {Natoli}, P. and {N{\o}rgaard-Nielsen}, H.~U. and {Pagano}, L. and {Paoletti}, D. and {Partridge}, B. and {Patanchon}, G. and {Peiris}, H.~V. and {Perrotta}, F. and {Pettorino}, V. and {Piacentini}, F. and {Polastri}, L. and {Polenta}, G. and {Puget}, J. -L. and {Rachen}, J.~P. and {Reinecke}, M. and {Remazeilles}, M. and {Renzi}, A. and {Rocha}, G. and {Rosset}, C. and {Roudier}, G. and {Rubi{\~n}o-Mart{\'\i}n}, J.~A. and {Ruiz-Granados}, B. and {Salvati}, L. and {Sandri}, M. and {Savelainen}, M. and {Scott}, D. and {Shellard}, E.~P.~S. and {Sirignano}, C. and {Sirri}, G. and {Spencer}, L.~D. and {Sunyaev}, R. and {Suur-Uski}, A. -S. and {Tauber}, J.~A. and {Tavagnacco}, D. and {Tenti}, M. and {Toffolatti}, L. and {Tomasi}, M. and {Trombetti}, T. and {Valenziano}, L. and {Valiviita}, J. and {Van Tent}, B. and {Vibert}, L. and {Vielva}, P. and {Villa}, F. and {Vittorio}, N. and {Wandelt}, B.~D. and {Wehus}, I.~K. and {White}, M. and {White}, S.~D.~M. and {Zacchei}, A. and {Zonca}, A.},
        title = "{Planck 2018 results. VI. Cosmological parameters}",
      journal = {\aap},
         year = 2020,
        month = sep,
       volume = {641},
          eid = {A6},
        pages = {A6},
          doi = {10.1051/0004-6361/201833910},
archivePrefix = {arXiv},
       eprint = {1807.06209},
 primaryClass = {astro-ph.CO},
       adsurl = {https://ui.adsabs.harvard.edu/abs/2020A&A...641A...6P}
}

@ARTICLE{PART2,
       author = {{DESI Collaboration} and {Aghamousa}, Amir and {Aguilar}, Jessica and {Ahlen}, Steve and {Alam}, Shadab and {Allen}, Lori E. and {Allende Prieto}, Carlos and {Annis}, James and {Bailey}, Stephen and {Balland}, Christophe and {Ballester}, Otger and {Baltay}, Charles and {Beaufore}, Lucas and {Bebek}, Chris and {Beers}, Timothy C. and {Bell}, Eric F. and {Bernal}, Jos{\'e} Luis and {Besuner}, Robert and {Beutler}, Florian and {Blake}, Chris and {Bleuler}, Hannes and {Blomqvist}, Michael and {Blum}, Robert and {Bolton}, Adam S. and {Briceno}, Cesar and {Brooks}, David and {Brownstein}, Joel R. and {Buckley-Geer}, Elizabeth and {Burden}, Angela and {Burtin}, Etienne and {Busca}, Nicolas G. and {Cahn}, Robert N. and {Cai}, Yan-Chuan and {Cardiel-Sas}, Laia and {Carlberg}, Raymond G. and {Carton}, Pierre-Henri and {Casas}, Ricard and {Castander}, Francisco J. and {Cervantes-Cota}, Jorge L. and {Claybaugh}, Todd M. and {Close}, Madeline and {Coker}, Carl T. and {Cole}, Shaun and {Comparat}, Johan and {Cooper}, Andrew P. and {Cousinou}, M. -C. and {Crocce}, Martin and {Cuby}, Jean-Gabriel and {Cunningham}, Daniel P. and {Davis}, Tamara M. and {Dawson}, Kyle S. and {de la Macorra}, Axel and {De Vicente}, Juan and {Delubac}, Timoth{\'e}e and {Derwent}, Mark and {Dey}, Arjun and {Dhungana}, Govinda and {Ding}, Zhejie and {Doel}, Peter and {Duan}, Yutong T. and {Ealet}, Anne and {Edelstein}, Jerry and {Eftekharzadeh}, Sarah and {Eisenstein}, Daniel J. and {Elliott}, Ann and {Escoffier}, St{\'e}phanie and {Evatt}, Matthew and {Fagrelius}, Parker and {Fan}, Xiaohui and {Fanning}, Kevin and {Farahi}, Arya and {Farihi}, Jay and {Favole}, Ginevra and {Feng}, Yu and {Fernandez}, Enrique and {Findlay}, Joseph R. and {Finkbeiner}, Douglas P. and {Fitzpatrick}, Michael J. and {Flaugher}, Brenna and {Flender}, Samuel and {Font-Ribera}, Andreu and {Forero-Romero}, Jaime E. and {Fosalba}, Pablo and {Frenk}, Carlos S. and {Fumagalli}, Michele and {Gaensicke}, Boris T. and {Gallo}, Giuseppe and {Garcia-Bellido}, Juan and {Gaztanaga}, Enrique and {Pietro Gentile Fusillo}, Nicola and {Gerard}, Terry and {Gershkovich}, Irena and {Giannantonio}, Tommaso and {Gillet}, Denis and {Gonzalez-de-Rivera}, Guillermo and {Gonzalez-Perez}, Violeta and {Gott}, Shelby and {Graur}, Or and {Gutierrez}, Gaston and {Guy}, Julien and {Habib}, Salman and {Heetderks}, Henry and {Heetderks}, Ian and {Heitmann}, Katrin and {Hellwing}, Wojciech A. and {Herrera}, David A. and {Ho}, Shirley and {Holland}, Stephen and {Honscheid}, Klaus and {Huff}, Eric and {Hutchinson}, Timothy A. and {Huterer}, Dragan and {Hwang}, Ho Seong and {Illa Laguna}, Joseph Maria and {Ishikawa}, Yuzo and {Jacobs}, Dianna and {Jeffrey}, Niall and {Jelinsky}, Patrick and {Jennings}, Elise and {Jiang}, Linhua and {Jimenez}, Jorge and {Johnson}, Jennifer and {Joyce}, Richard and {Jullo}, Eric and {Juneau}, St{\'e}phanie and {Kama}, Sami and {Karcher}, Armin and {Karkar}, Sonia and {Kehoe}, Robert and {Kennamer}, Noble and {Kent}, Stephen and {Kilbinger}, Martin and {Kim}, Alex G. and {Kirkby}, David and {Kisner}, Theodore and {Kitanidis}, Ellie and {Kneib}, Jean-Paul and {Koposov}, Sergey and {Kovacs}, Eve and {Koyama}, Kazuya and {Kremin}, Anthony and {Kron}, Richard and {Kronig}, Luzius and {Kueter-Young}, Andrea and {Lacey}, Cedric G. and {Lafever}, Robin and {Lahav}, Ofer and {Lambert}, Andrew and {Lampton}, Michael and {Landriau}, Martin and {Lang}, Dustin and {Lauer}, Tod R. and {Le Goff}, Jean-Marc and {Le Guillou}, Laurent and {Le Van Suu}, Auguste and {Lee}, Jae Hyeon and {Lee}, Su-Jeong and {Leitner}, Daniela and {Lesser}, Michael and {Levi}, Michael E. and {L'Huillier}, Benjamin and {Li}, Baojiu and {Liang}, Ming and {Lin}, Huan and {Linder}, Eric and {Loebman}, Sarah R. and {Luki{\'c}}, Zarija and {Ma}, Jun and {MacCrann}, Niall and {Magneville}, Christophe and {Makarem}, Laleh and {Manera}, Marc and {Manser}, Christopher J. and {Marshall}, Robert and {Martini}, Paul and {Massey}, Richard and {Matheson}, Thomas and {McCauley}, Jeremy and {McDonald}, Patrick and {McGreer}, Ian D. and {Meisner}, Aaron and {Metcalfe}, Nigel and {Miller}, Timothy N. and {Miquel}, Ramon and {Moustakas}, John and {Myers}, Adam and {Naik}, Milind and {Newman}, Jeffrey A. and {Nichol}, Robert C. and {Nicola}, Andrina and {Nicolati da Costa}, Luiz and {Nie}, Jundan and {Niz}, Gustavo and {Norberg}, Peder and {Nord}, Brian and {Norman}, Dara and {Nugent}, Peter and {O'Brien}, Thomas and {Oh}, Minji and {Olsen}, Knut A.~G.},
        title = "{The DESI Experiment Part II: Instrument Design}",
      journal = {arXiv e-prints},
         year = 2016,
        month = oct,
          eid = {arXiv:1611.00037},
        pages = {arXiv:1611.00037},
          doi = {10.48550/arXiv.1611.00037},
archivePrefix = {arXiv},
       eprint = {1611.00037},
 primaryClass = {astro-ph.IM},
       adsurl = {https://ui.adsabs.harvard.edu/abs/2016arXiv161100037D}
}

@ARTICLE{TARGET_2N,
       author = {{Myers}, Adam D. and {Moustakas}, John and {Bailey}, Stephen and {Weaver}, Benjamin A. and {Cooper}, Andrew P. and {Forero-Romero}, Jaime E. and {Abolfathi}, Bela and {Alexander}, David M. and {Brooks}, David and {Chaussidon}, Edmond and {Chuang}, Chia-Hsun and {Dawson}, Kyle and {Dey}, Arjun and {Dey}, Biprateep and {Dhungana}, Govinda and {Doel}, Peter and {Fanning}, Kevin and {Gazta{\~n}aga}, Enrique and {Gontcho A Gontcho}, Satya and {Gonzalez-Morales}, Alma X. and {Hahn}, ChangHoon and {Herrera-Alcantar}, Hiram K. and {Honscheid}, Klaus and {Ishak}, Mustapha and {Karim}, Tanveer and {Kirkby}, David and {Kisner}, Theodore and {Koposov}, Sergey E. and {Kremin}, Anthony and {Lan}, Ting-Wen and {Landriau}, Martin and {Lang}, Dustin and {Levi}, Michael E. and {Magneville}, Christophe and {Napolitano}, Lucas and {Martini}, Paul and {Meisner}, Aaron and {Newman}, Jeffrey A. and {Palanque-Delabrouille}, Nathalie and {Percival}, Will and {Poppett}, Claire and {Prada}, Francisco and {Raichoor}, Anand and {Ross}, Ashley J. and {Schlafly}, Edward F. and {Schlegel}, David and {Schubnell}, Michael and {Tan}, Ting and {Tarle}, Gregory and {Wilson}, Michael J. and {Y{\`e}che}, Christophe and {Zhou}, Rongpu and {Zhou}, Zhimin and {Zou}, Hu},
        title = "{The Target-selection Pipeline for the Dark Energy Spectroscopic Instrument}",
      journal = {\aj},
         year = 2023,
        month = feb,
       volume = {165},
       number = {2},
          eid = {50},
        pages = {50},
          doi = {10.3847/1538-3881/aca5f9},
archivePrefix = {arXiv},
       eprint = {2208.08518},
 primaryClass = {astro-ph.IM},
       adsurl = {https://ui.adsabs.harvard.edu/abs/2023AJ....165...50M}
}

@ARTICLE{BCG_catalog,
       author = {{Zou}, Hu and {Sui}, Jipeng and {Xue}, Suijian and {Zhou}, Xu and {Ma}, Jun and {Zhou}, Zhimin and {Nie}, Jundan and {Zhang}, Tianmeng and {Feng}, Lu and {Shen}, Zhixia and {Wang}, Jiali},
        title = "{Photometric Redshifts and Galaxy Clusters for DES DR2, DESI DR9, and HSC-SSP PDR3 Data}",
      journal = {Research in Astronomy and Astrophysics},
         year = 2022,
        month = jun,
       volume = {22},
       number = {6},
          eid = {065001},
        pages = {065001},
          doi = {10.1088/1674-4527/ac6416},
archivePrefix = {arXiv},
       eprint = {2203.17035},
 primaryClass = {astro-ph.GA},
       adsurl = {https://ui.adsabs.harvard.edu/abs/2022RAA....22f5001Z}
}

@software{QSOfit1,
       author = {{Guo}, Hengxiao and {Shen}, Yue and {Wang}, Shu},
        title = "{PyQSOFit: Python code to fit the spectrum of quasars}",
 howpublished = {Astrophysics Source Code Library, record ascl:1809.008},
         year = 2018,
        month = sep,
          eid = {ascl:1809.008},
       adsurl = {https://ui.adsabs.harvard.edu/abs/2018ascl.soft09008G}
}

@ARTICLE{QSOfit2,
       author = {{Shen}, Yue and {Hall}, Patrick B. and {Horne}, Keith and {Zhu}, Guangtun and {McGreer}, Ian and {Simm}, Torben and {Trump}, Jonathan R. and {Kinemuchi}, Karen and {Brandt}, W.~N. and {Green}, Paul J. and {Grier}, C.~J. and {Guo}, Hengxiao and {Ho}, Luis C. and {Homayouni}, Yasaman and {Jiang}, Linhua and {I-Hsiu Li}, Jennifer and {Morganson}, Eric and {Petitjean}, Patrick and {Richards}, Gordon T. and {Schneider}, Donald P. and {Starkey}, D.~A. and {Wang}, Shu and {Chambers}, Ken and {Kaiser}, Nick and {Kudritzki}, Rolf-Peter and {Magnier}, Eugene and {Waters}, Christopher},
        title = "{The Sloan Digital Sky Survey Reverberation Mapping Project: Sample Characterization}",
      journal = {\apjs},
         year = 2019,
        month = apr,
       volume = {241},
       number = {2},
          eid = {34},
        pages = {34},
          doi = {10.3847/1538-4365/ab074f},
archivePrefix = {arXiv},
       eprint = {1810.01447},
 primaryClass = {astro-ph.GA},
       adsurl = {https://ui.adsabs.harvard.edu/abs/2019ApJS..241...34S}
}

@ARTICLE{2025ApJ...981...81C,
       author = {{Chen}, Zeyu and {Wang}, Enci and {Zou}, Hu and {Zou}, Siwei and {Gao}, Yang and {Wang}, Huiyuan and {Yu}, Haoran and {Jia}, Cheng and {Li}, Haixin and {Ma}, Chengyu and {Yao}, Yao and {Ding}, Weiyu and {Zhu}, Runyu},
        title = "{The Circumgalactic Medium Traced by Mg II Absorption with DESI: Dependence on Galaxy Stellar Mass, Star Formation Rate, and Azimuthal Angle}",
      journal = {\apj},
         year = 2025,
        month = mar,
       volume = {981},
       number = {1},
          eid = {81},
        pages = {81},
          doi = {10.3847/1538-4357/ada942},
archivePrefix = {arXiv},
       eprint = {2411.08485},
 primaryClass = {astro-ph.GA},
       adsurl = {https://ui.adsabs.harvard.edu/abs/2025ApJ...981...81C}
}

@ARTICLE{2013ApJ...770..130Z,
       author = {{Zhu}, Guangtun and {M{\'e}nard}, Brice},
        title = "{The JHU-SDSS Metal Absorption Line Catalog: Redshift Evolution and Properties of Mg II Absorbers}",
      journal = {\apj},
         year = 2013,
        month = jun,
       volume = {770},
       number = {2},
          eid = {130},
        pages = {130},
          doi = {10.1088/0004-637X/770/2/130},
archivePrefix = {arXiv},
       eprint = {1211.6215},
 primaryClass = {astro-ph.CO},
       adsurl = {https://ui.adsabs.harvard.edu/abs/2013ApJ...770..130Z}
}

@ARTICLE{2013PASP..125..306F,
       author = {{Foreman-Mackey}, Daniel and {Hogg}, David W. and {Lang}, Dustin and {Goodman}, Jonathan},
        title = "{emcee: The MCMC Hammer}",
      journal = {\pasp},
         year = 2013,
        month = mar,
       volume = {125},
       number = {925},
        pages = {306},
          doi = {10.1086/670067},
archivePrefix = {arXiv},
       eprint = {1202.3665},
 primaryClass = {astro-ph.IM},
       adsurl = {https://ui.adsabs.harvard.edu/abs/2013PASP..125..306F}
}

@ARTICLE{Lan2018ApJ,
       author = {{Lan}, Ting-Wen and {Mo}, Houjun},
        title = "{The Circumgalactic Medium of eBOSS Emission Line Galaxies: Signatures of Galactic Outflows in Gas Distribution and Kinematics}",
      journal = {\apj},
         year = 2018,
        month = oct,
       volume = {866},
       number = {1},
          eid = {36},
        pages = {36},
          doi = {10.3847/1538-4357/aadc08},
archivePrefix = {arXiv},
       eprint = {1806.05786},
 primaryClass = {astro-ph.GA},
       adsurl = {https://ui.adsabs.harvard.edu/abs/2018ApJ...866...36L}
}

@ARTICLE{Anand_2021,
       author = {{Anand}, Abhijeet and {Nelson}, Dylan and {Kauffmann}, Guinevere},
        title = "{Characterizing the abundance, properties, and kinematics of the cool circumgalactic medium of galaxies in absorption with SDSS DR16}",
      journal = {\mnras},
         year = 2021,
        month = jun,
       volume = {504},
       number = {1},
        pages = {65-88},
          doi = {10.1093/mnras/stab871},
archivePrefix = {arXiv},
       eprint = {2103.15842},
 primaryClass = {astro-ph.GA},
       adsurl = {https://ui.adsabs.harvard.edu/abs/2021MNRAS.504...65A}
}

@ARTICLE{Anand_2022,
       author = {{Anand}, Abhijeet and {Kauffmann}, Guinevere and {Nelson}, Dylan},
        title = "{Cool circumgalactic gas in galaxy clusters: connecting the DESI legacy imaging survey and SDSS DR16 Mg II absorbers}",
      journal = {\mnras},
         year = 2022,
        month = jul,
       volume = {513},
       number = {3},
        pages = {3210-3227},
          doi = {10.1093/mnras/stac928},
archivePrefix = {arXiv},
       eprint = {2201.07811},
 primaryClass = {astro-ph.GA},
       adsurl = {https://ui.adsabs.harvard.edu/abs/2022MNRAS.513.3210A}
}

@ARTICLE{DESI_DR1,
       author = {{DESI Collaboration} and {Abdul Karim}, M. and {Adame}, A.~G. and {Aguado}, D. and {Aguilar}, J. and {Ahlen}, S. and {Alam}, S. and {Aldering}, G. and {Alexander}, D.~M. and {Alfarsy}, R. and {Allen}, L. and {Allende Prieto}, C. and {Alves}, O. and {Anand}, A. and {Andrade}, U. and {Armengaud}, E. and {Avila}, S. and {Aviles}, A. and {Awan}, H. and {Bailey}, S. and {Baleato Lizancos}, A. and {Ballester}, O. and {Bault}, A. and {Bautista}, J. and {Bean}, R. and {Behera}, J. and {BenZvi}, S. and {Beraldo e Silva}, L. and {Bermejo-Climent}, J.~R. and {Beutler}, F. and {Bianchi}, D. and {Blake}, C. and {Blum}, R. and {Bolton}, A.~S. and {Bonici}, M. and {Brieden}, S. and {Brodzeller}, A. and {Brooks}, D. and {Buckley-Geer}, E. and {Burtin}, E. and {Bystr{\"o}m}, A. and {Canning}, R. and {Carnero Rosell}, A. and {Carr}, A. and {Carrilho}, P. and {Casas}, L. and {Castander}, F.~J. and {Cereskaite}, R. and {Cervantes-Cota}, J.~L. and {Chaussidon}, E. and {Chaves-Montero}, J. and {Chen}, S. and {Chen}, X. and {Circosta}, C. and {Claybaugh}, T. and {Cole}, S. and {Cooper}, A.~P. and {Cousinou}, M.-C. and {Cuceu}, A. and {Davis}, T.~M. and {Dawson}, K.~S. and {de Belsunce}, R. and {de la Cruz}, R. and {de la Macorra}, A. and {de Mattia}, A. and {Deiosso}, N. and {Della Costa}, J. and {Demina}, R. and {Demirbozan}, U. and {DeRose}, J. and {Dey}, A. and {Dey}, B. and {Ding}, J. and {Ding}, Z. and {Doel}, P. and {Douglass}, K. and {Dowicz}, M. and {Ebina}, H. and {Edelstein}, J. and {Eisenstein}, D.~J. and {Elbers}, W. and {Emas}, N. and {Escoffier}, S. and {Fagrelius}, P. and {Fan}, X. and {Fanning}, K. and {Favole}, G. and {Fawcett}, V.~A. and {Fern{\'a}ndez-Garc{\'\i}a}, E. and {Ferraro}, S. and {Findlay}, N. and {Font-Ribera}, A. and {Forero-Romero}, J.~E. and {Forero-S{\'a}nchez}, D. and {Frenk}, C.~S. and {G{\"a}nsicke}, B.~T. and {Galbany}, L. and {Garc{\'\i}a-Bellido}, J. and {Garcia-Quintero}, C. and {Garrison}, L.~H. and {Gazta{\~n}aga}, E. and {Gil-Mar{\'\i}n}, H. and {Gloudemans}, A. and {Gnedin}, O.~Y. and {Gontcho A Gontcho}, S. and {Gonzalez}, D. and {Gonzalez-Morales}, A.~X. and {Gonzalez-Perez}, V. and {Gordon}, C. and {Graur}, O. and {Green}, D. and {Gruen}, D. and {Gsponer}, R. and {Guandalin}, C. and {Gutierrez}, G. and {Guy}, J. and {Hahn}, C. and {Han}, J.~J. and {Han}, J. and {He}, S. and {Herrera-Alcantar}, H.~K. and {Heydenreich}, S. and {Honscheid}, K. and {Hou}, J. and {Howlett}, C. and {Huterer}, D. and {Ir{\v{s}}i{\v{c}}}, V. and {Ishak}, M. and {Jacques}, A. and {Jiang}, L. and {Jimenez}, J. and {Jing}, Y.~P. and {Joachimi}, B. and {Joudaki}, S. and {Joyce}, R. and {Jullo}, E. and {Juneau}, S. and {Kara{\c{c}}ayl{\i}}, N.~G. and {Karim}, T. and {Kehoe}, R. and {Kent}, S. and {Khederlarian}, A. and {Kirkby}, D. and {Kisner}, T. and {Kitaura}, F.-S. and {Kizhuprakkat}, N. and {Kong}, H. and {Koposov}, S.~E. and {Kremin}, A. and {Krolewski}, A. and {Lahav}, O. and {Lai}, Y. and {Lamman}, C. and {Lan}, T.-W. and {Landriau}, M. and {Lang}, D. and {Lange}, J.~U. and {Lasker}, J. and {Le Goff}, J.~M. and {Le Guillou}, L. and {Leauthaud}, A. and {Levi}, M.~E. and {Li}, S. and {Li}, T.~S. and {Liu}, W. and {Lodha}, K. and {Lokken}, M. and {Luo}, Y. and {Magneville}, C. and {Manera}, M. and {Manser}, C.~J. and {Margala}, D. and {Martini}, P. and {Maus}, M. and {McCullough}, J. and {McDonald}, P. and {Medina}, G.~E. and {Medina-Varela}, L. and {Meisner}, A. and {Mena-Fern{\'a}ndez}, J. and {Menegas}, A. and {Meneses-Rizo}, J. and {Mezcua}, M. and {Miquel}, R. and {Montero-Camacho}, P. and {Moon}, J. and {Moustakas}, J. and {Mu{\~n}oz-Guti{\'e}rrez}, A. and {Mu noz-Santos}, D. and {Myers}, A.~D. and {Myles}, J. and {Nadathur}, S. and {Najita}, J. and {Napolitano}, L. and {Newman}, J.~A. and {Nikakhtar}, F. and {Nikutta}, R. and {Niz}, G. and {Noriega}, H.~E. and {Nugent}, P.},
        title = "{Data Release 1 of the Dark Energy Spectroscopic Instrument}",
      journal = {\aj},
         year = 2026,
        month = may,
       volume = {171},
       number = {5},
          eid = {285},
        pages = {285},
          doi = {10.3847/1538-3881/ae4c43},
archivePrefix = {arXiv},
       eprint = {2503.14745},
 primaryClass = {astro-ph.CO},
       adsurl = {https://ui.adsabs.harvard.edu/abs/2026AJ....171..285D}
}

@ARTICLE{Tumlinson_2017ARA&A..55..389T,
       author = {{Tumlinson}, Jason and {Peeples}, Molly S. and {Werk}, Jessica K.},
        title = "{The Circumgalactic Medium}",
      journal = {\araa},
         year = 2017,
        month = aug,
       volume = {55},
       number = {1},
        pages = {389-432},
          doi = {10.1146/annurev-astro-091916-055240},
archivePrefix = {arXiv},
       eprint = {1709.09180},
 primaryClass = {astro-ph.GA},
       adsurl = {https://ui.adsabs.harvard.edu/abs/2017ARA&A..55..389T}
}

@ARTICLE{2020ARA&A..58..363P,
       author = {{P{\'e}roux}, C{\'e}line and {Howk}, J. Christopher},
        title = "{The Cosmic Baryon and Metal Cycles}",
      journal = {\araa},
         year = 2020,
        month = aug,
       volume = {58},
        pages = {363-406},
          doi = {10.1146/annurev-astro-021820-120014},
archivePrefix = {arXiv},
       eprint = {2011.01935},
 primaryClass = {astro-ph.GA},
       adsurl = {https://ui.adsabs.harvard.edu/abs/2020ARA&A..58..363P}
}

@ARTICLE{Weinberger_2017MNRAS.465.3291W,
       author = {{Weinberger}, Rainer and {Springel}, Volker and {Hernquist}, Lars and {Pillepich}, Annalisa and {Marinacci}, Federico and {Pakmor}, R{\"u}diger and {Nelson}, Dylan and {Genel}, Shy and {Vogelsberger}, Mark and {Naiman}, Jill and {Torrey}, Paul},
        title = "{Simulating galaxy formation with black hole driven thermal and kinetic feedback}",
      journal = {\mnras},
         year = 2017,
        month = mar,
       volume = {465},
       number = {3},
        pages = {3291-3308},
          doi = {10.1093/mnras/stw2944},
archivePrefix = {arXiv},
       eprint = {1607.03486},
 primaryClass = {astro-ph.GA},
       adsurl = {https://ui.adsabs.harvard.edu/abs/2017MNRAS.465.3291W}
}

@ARTICLE{Burkhart_2022ApJ...933L..46B,
       author = {{Burkhart}, Blakesley and {Tillman}, Megan and {Gurvich}, Alexander B. and {Bird}, Simeon and {Tonnesen}, Stephanie and {Bryan}, Greg L. and {Hernquist}, Lars and {Somerville}, Rachel S.},
        title = "{The Low-redshift Ly{\ensuremath{\alpha}} Forest as a Constraint for Models of AGN Feedback}",
      journal = {\apjl},
         year = 2022,
        month = jul,
       volume = {933},
       number = {2},
          eid = {L46},
        pages = {L46},
          doi = {10.3847/2041-8213/ac7e49},
archivePrefix = {arXiv},
       eprint = {2204.09712},
 primaryClass = {astro-ph.GA},
       adsurl = {https://ui.adsabs.harvard.edu/abs/2022ApJ...933L..46B}
}

@ARTICLE{Lundgren_2009ApJ...698..819L,
       author = {{Lundgren}, Britt F. and {Brunner}, Robert J. and {York}, Donald G. and {Ross}, Ashley J. and {Quashnock}, Jean M. and {Myers}, Adam D. and {Schneider}, Donald P. and {Al Sayyad}, Yusra and {Bahcall}, Neta},
        title = "{A Cross-Correlation Analysis of Mg II Absorption Line Systems and Luminous Red Galaxies from the SDSS DR5}",
      journal = {\apj},
         year = 2009,
        month = jun,
       volume = {698},
       number = {1},
        pages = {819-839},
          doi = {10.1088/0004-637X/698/1/819},
archivePrefix = {arXiv},
       eprint = {0902.4003},
 primaryClass = {astro-ph.CO},
       adsurl = {https://ui.adsabs.harvard.edu/abs/2009ApJ...698..819L}
}

@ARTICLE{Huang_2016MNRAS.455.1713H,
       author = {{Huang}, Yun-Hsin and {Chen}, Hsiao-Wen and {Johnson}, Sean D. and {Weiner}, Benjamin J.},
        title = "{Characterizing the chemically enriched circumgalactic medium of {\ensuremath{\sim}}38 000 luminous red galaxies in SDSS DR12}",
      journal = {\mnras},
         year = 2016,
        month = jan,
       volume = {455},
       number = {2},
        pages = {1713-1727},
          doi = {10.1093/mnras/stv2327},
archivePrefix = {arXiv},
       eprint = {1510.01336},
 primaryClass = {astro-ph.GA},
       adsurl = {https://ui.adsabs.harvard.edu/abs/2016MNRAS.455.1713H}
}

@ARTICLE{Smailagic_2018ApJ...867..106S,
       author = {{Smailagi{\'c}}, Marijana and {Prochaska}, Jason Xavier and {Burchett}, Joseph and {Zhu}, Guangtun and {M{\'e}nard}, Brice},
        title = "{Extreme Circumgalactic H I and C III Absorption around the Most Massive, Quenched Galaxies}",
      journal = {\apj},
         year = 2018,
        month = nov,
       volume = {867},
       number = {2},
          eid = {106},
        pages = {106},
          doi = {10.3847/1538-4357/aae384},
archivePrefix = {arXiv},
       eprint = {1809.09777},
 primaryClass = {astro-ph.GA},
       adsurl = {https://ui.adsabs.harvard.edu/abs/2018ApJ...867..106S}
}

@ARTICLE{Zahedy_2019MNRAS.484.2257Z,
       author = {{Zahedy}, Fakhri S. and {Chen}, Hsiao-Wen and {Johnson}, Sean D. and {Pierce}, Rebecca M. and {Rauch}, Michael and {Huang}, Yun-Hsin and {Weiner}, Benjamin J. and {Gauthier}, Jean-Ren{\'e}},
        title = "{Characterizing circumgalactic gas around massive ellipticals at z {\ensuremath{\sim}} 0.4 - II. Physical properties and elemental abundances}",
      journal = {\mnras},
         year = 2019,
        month = apr,
       volume = {484},
       number = {2},
        pages = {2257-2280},
          doi = {10.1093/mnras/sty3482},
archivePrefix = {arXiv},
       eprint = {1809.05115},
 primaryClass = {astro-ph.GA},
       adsurl = {https://ui.adsabs.harvard.edu/abs/2019MNRAS.484.2257Z}
}

@ARTICLE{2014MNRAS.439.3139Z,
       author = {{Zhu}, Guangtun and {M{\'e}nard}, Brice and {Bizyaev}, Dmitry and {Brewington}, Howard and {Ebelke}, Garrett and {Ho}, Shirley and {Kinemuchi}, Karen and {Malanushenko}, Viktor and {Malanushenko}, Elena and {Marchante}, Moses and {More}, Surhud and {Oravetz}, Daniel and {Pan}, Kaike and {Petitjean}, Patrick and {Simmons}, Audrey},
        title = "{The large-scale distribution of cool gas around luminous red galaxies}",
      journal = {\mnras},
         year = 2014,
        month = apr,
       volume = {439},
       number = {3},
        pages = {3139-3155},
          doi = {10.1093/mnras/stu186},
archivePrefix = {arXiv},
       eprint = {1309.7660},
 primaryClass = {astro-ph.CO},
       adsurl = {https://ui.adsabs.harvard.edu/abs/2014MNRAS.439.3139Z}
}

@ARTICLE{Wu_2025ApJ...983..186W,
       author = {{Wu}, Xuanyi and {Cai}, Z. and {Lan}, T. -W. and {Zou}, S. and {Anand}, A. and {Dey}, Biprateep and {Li}, Z. and {Aguilar}, J. and {Ahlen}, S. and {Brooks}, D. and {Claybaugh}, T. and {de la Macorra}, A. and {Doel}, P. and {Ferraro}, S. and {Forero-Romero}, J.~E. and {Gontcho A Gontcho}, S. and {Honscheid}, K. and {Juneau}, S. and {Kehoe}, R. and {Kisner}, T. and {Lambert}, A. and {Landriau}, M. and {Le Guillou}, L. and {Manera}, M. and {Meisner}, A. and {Miquel}, R. and {Moustakas}, J. and {Newman}, J.~A. and {Prada}, F. and {Rossi}, G. and {Sanchez}, E. and {Schlegel}, D. and {Schubnell}, M. and {Siudek}, M. and {Sprayberry}, D. and {Tarl{\'e}}, G. and {Weaver}, B.~A. and {Zou}, H.},
        title = "{Tracing the Evolution of the Cool Gas in CGM and IGM Environments through Mg II Absorption from Redshift z = 0.75 to z = 1.65 Using DESI-Y1 Data}",
      journal = {\apj},
         year = 2025,
        month = apr,
       volume = {983},
       number = {2},
          eid = {186},
        pages = {186},
          doi = {10.3847/1538-4357/adb28a},
archivePrefix = {arXiv},
       eprint = {2407.17809},
 primaryClass = {astro-ph.GA},
       adsurl = {https://ui.adsabs.harvard.edu/abs/2025ApJ...983..186W}
}

@ARTICLE{2024MNRAS.527.1301R,
       author = {{Rost}, Agust{\'\i}n M. and {Nuza}, Sebasti{\'a}n E. and {Stasyszyn}, Federico and {Kuchner}, Ulrike and {Hoeft}, Matthias and {Welker}, Charlotte and {Pearce}, Frazer and {Gray}, Meghan and {Knebe}, Alexander and {Cui}, Weiguang and {Yepes}, Gustavo},
        title = "{The three hundred project: thermodynamical properties, shocks, and gas dynamics in simulated galaxy cluster filaments and their surroundings}",
      journal = {\mnras},
         year = 2024,
        month = jan,
       volume = {527},
       number = {1},
        pages = {1301-1316},
          doi = {10.1093/mnras/stad3208},
archivePrefix = {arXiv},
       eprint = {2310.12245},
 primaryClass = {astro-ph.CO},
       adsurl = {https://ui.adsabs.harvard.edu/abs/2024MNRAS.527.1301R}
}

@ARTICLE{2025arXiv250301960S,
       author = {{Staffehl}, Milan and {Nelson}, Dylan and {Ayromlou}, Mohammadreza and {Rohr}, Eric and {Pillepich}, Annalisa},
        title = "{The abundance and origin of cool gas in galaxy clusters in the TNG-Cluster simulation}",
      journal = {arXiv e-prints},
         year = 2025,
        month = mar,
          eid = {arXiv:2503.01960},
        pages = {arXiv:2503.01960},
          doi = {10.48550/arXiv.2503.01960},
archivePrefix = {arXiv},
       eprint = {2503.01960},
 primaryClass = {astro-ph.GA},
       adsurl = {https://ui.adsabs.harvard.edu/abs/2025arXiv250301960S}
}

@ARTICLE{2022A&ARv..30....3B,
       author = {{Boselli}, Alessandro and {Fossati}, Matteo and {Sun}, Ming},
        title = "{Ram pressure stripping in high-density environments}",
      journal = {\aapr},
         year = 2022,
        month = dec,
       volume = {30},
       number = {1},
          eid = {3},
        pages = {3},
          doi = {10.1007/s00159-022-00140-3},
archivePrefix = {arXiv},
       eprint = {2109.13614},
 primaryClass = {astro-ph.GA},
       adsurl = {https://ui.adsabs.harvard.edu/abs/2022A&ARv..30....3B}
}

@ARTICLE{2007MNRAS.382.1481N,
       author = {{Nipoti}, Carlo and {Binney}, James},
        title = "{The role of thermal evaporation in galaxy formation}",
      journal = {\mnras},
         year = 2007,
        month = dec,
       volume = {382},
       number = {4},
        pages = {1481-1493},
          doi = {10.1111/j.1365-2966.2007.12505.x},
archivePrefix = {arXiv},
       eprint = {0707.4147},
 primaryClass = {astro-ph},
       adsurl = {https://ui.adsabs.harvard.edu/abs/2007MNRAS.382.1481N}
}

@ARTICLE{1972ApJ...176....1G,
       author = {{Gunn}, James E. and {Gott}, III, J. Richard},
        title = "{On the Infall of Matter Into Clusters of Galaxies and Some Effects on Their Evolution}",
      journal = {\apj},
         year = 1972,
        month = aug,
       volume = {176},
        pages = {1},
          doi = {10.1086/151605},
       adsurl = {https://ui.adsabs.harvard.edu/abs/1972ApJ...176....1G}
}

@ARTICLE{1990ApJ...350...89B,
       author = {{Byrd}, Gene and {Valtonen}, Mauri},
        title = "{Tidal Generation of Active Spirals and S0 Galaxies by Rich Clusters}",
      journal = {\apj},
         year = 1990,
        month = feb,
       volume = {350},
        pages = {89},
          doi = {10.1086/168362},
       adsurl = {https://ui.adsabs.harvard.edu/abs/1990ApJ...350...89B}
}

@ARTICLE{1980ApJ...237..692L,
       author = {{Larson}, R.~B. and {Tinsley}, B.~M. and {Caldwell}, C.~N.},
        title = "{The evolution of disk galaxies and the origin of S0 galaxies}",
      journal = {\apj},
         year = 1980,
        month = may,
       volume = {237},
        pages = {692-707},
          doi = {10.1086/157917},
       adsurl = {https://ui.adsabs.harvard.edu/abs/1980ApJ...237..692L}
}

@ARTICLE{1984ApJ...281...95P,
       author = {{Postman}, M. and {Geller}, M.~J.},
        title = "{The morphology-density relation - The group connection.}",
      journal = {\apj},
         year = 1984,
        month = jun,
       volume = {281},
        pages = {95-99},
          doi = {10.1086/162078},
       adsurl = {https://ui.adsabs.harvard.edu/abs/1984ApJ...281...95P}
}

@ARTICLE{2010ApJ...721..193P,
       author = {{Peng}, Ying-jie and {Lilly}, Simon J. and {Kova{\v{c}}}, Katarina and {Bolzonella}, Micol and {Pozzetti}, Lucia and {Renzini}, Alvio and {Zamorani}, Gianni and {Ilbert}, Olivier and {Knobel}, Christian and {Iovino}, Angela and {Maier}, Christian and {Cucciati}, Olga and {Tasca}, Lidia and {Carollo}, C. Marcella and {Silverman}, John and {Kampczyk}, Pawel and {de Ravel}, Loic and {Sanders}, David and {Scoville}, Nicholas and {Contini}, Thierry and {Mainieri}, Vincenzo and {Scodeggio}, Marco and {Kneib}, Jean-Paul and {Le F{\`e}vre}, Olivier and {Bardelli}, Sandro and {Bongiorno}, Angela and {Caputi}, Karina and {Coppa}, Graziano and {de la Torre}, Sylvain and {Franzetti}, Paolo and {Garilli}, Bianca and {Lamareille}, Fabrice and {Le Borgne}, Jean-Francois and {Le Brun}, Vincent and {Mignoli}, Marco and {Perez Montero}, Enrique and {Pello}, Roser and {Ricciardelli}, Elena and {Tanaka}, Masayuki and {Tresse}, Laurence and {Vergani}, Daniela and {Welikala}, Niraj and {Zucca}, Elena and {Oesch}, Pascal and {Abbas}, Ummi and {Barnes}, Luke and {Bordoloi}, Rongmon and {Bottini}, Dario and {Cappi}, Alberto and {Cassata}, Paolo and {Cimatti}, Andrea and {Fumana}, Marco and {Hasinger}, Gunther and {Koekemoer}, Anton and {Leauthaud}, Alexei and {Maccagni}, Dario and {Marinoni}, Christian and {McCracken}, Henry and {Memeo}, Pierdomenico and {Meneux}, Baptiste and {Nair}, Preethi and {Porciani}, Cristiano and {Presotto}, Valentina and {Scaramella}, Roberto},
        title = "{Mass and Environment as Drivers of Galaxy Evolution in SDSS and zCOSMOS and the Origin of the Schechter Function}",
      journal = {\apj},
         year = 2010,
        month = sep,
       volume = {721},
       number = {1},
        pages = {193-221},
          doi = {10.1088/0004-637X/721/1/193},
archivePrefix = {arXiv},
       eprint = {1003.4747},
 primaryClass = {astro-ph.CO},
       adsurl = {https://ui.adsabs.harvard.edu/abs/2010ApJ...721..193P}
}
\bibliographystyle{aasjournal}



\end{document}